\documentclass[acmsmall,nonacm]{acmart}
\pdfoutput=1

\usepackage{type1cm}        
\usepackage{graphicx}        
\usepackage{multicol}        
\usepackage[bottom]{footmisc}

\usepackage{newtxmath}       

\usepackage{longtable}
\usepackage{ltcaption}
\usepackage{enumitem}
\usepackage{float}

\usepackage{pifont}
\newcommand{\cmark}{\ding{51}}%
\newcommand{\xmark}{\ding{55}}%

\usepackage{color}
\definecolor{lightgray}{gray}{0.85}

\usepackage{xcolor}
\definecolor{Blue}{rgb}{0,0,1}

\definecolor{Orange}{rgb}{1,0.5,0}

\definecolor{Green}{rgb}{0,1,0}

\usepackage{comment}

\begin{document}

\title{Algorithmic Fairness}

\author{Dana Pessach}
\authornote{Corresponding author}
\affiliation{%
  \institution{Department of Industrial Engineering, Tel-Aviv University}
  \streetaddress{P.O. Box 39040}
  \postcode{6997801}
  \city{Tel-Aviv}
  \country{Israel}}
\email{danapessach@gmail.com}

\author{Erez Shmueli}
\affiliation{%
  \institution{Department of Industrial Engineering, Tel-Aviv University}
  \streetaddress{P.O. Box 39040}
  \postcode{6997801}
  \city{Tel-Aviv}
  \country{Israel}}
\email{shmueli@tau.ac.il}

\begin{abstract}
An increasing number of decisions regarding the daily lives of human beings are being controlled by artificial intelligence (AI) algorithms in spheres ranging from healthcare, transportation, and education to college admissions, recruitment, provision of loans and many more realms.
Since they now touch on many aspects of our lives, it is crucial to develop AI algorithms that are not only accurate but also objective and fair.
Recent studies have shown that algorithmic decision-making may be inherently prone to unfairness, even when there is no intention for it.
This paper presents an overview of the main concepts of identifying, measuring and improving algorithmic fairness when using AI algorithms.
The paper begins by discussing the causes of algorithmic bias and unfairness and the common definitions and measures for fairness.
Fairness-enhancing mechanisms are then reviewed and divided into pre-process, in-process and post-process mechanisms.
A comprehensive comparison of the mechanisms is then conducted, towards a better understanding of which mechanisms should be used in different scenarios.
The paper then describes the most commonly used fairness-related datasets in this field.
Finally, the paper ends by reviewing several emerging research sub-fields of algorithmic fairness.


\end{abstract}

\maketitle

\section{Introduction}
\label{S:1}

Nowadays, an increasing number of decisions are being controlled by artificial intelligence (AI) algorithms, with increased implementation of automated decision-making systems in business and government applications.
The motivation for an automated learning model is clear -- we expect algorithms to perform better than human beings for several reasons: First, algorithms may integrate much more data than a human may grasp and take many more considerations into account.
Second, algorithms can perform complex computations much faster than human beings.
Third, human decisions are subjective, and they often include biases.

Hence, it is a common belief that using an automated algorithm makes decisions more objective or fair.
However, this is unfortunately not the case since AI algorithms are not always as objective as we would expect.
The idea that AI algorithms are free from biases is wrong since the assumption that the data injected into the models are unbiased is wrong.
More specifically, a prediction model may actually be inherently biased since it learns and preserves historical biases \citep{kleinberg2017inherent}.

Since many automated decisions (including which individuals will receive jobs, loans, medication, bail or parole) can significantly impact people's lives, there is great importance in assessing and improving the ethics of the decisions made by these automated systems.
Indeed, in recent years, the concern for algorithm fairness has made headlines.
One of the most common examples was in the field of criminal justice, where recent revelations have shown that an algorithm used by the United States criminal justice system had falsely predicted future criminality among African-Americans at twice the rate as it predicted for white people \citep{Angwin:2016:Online, chouldechova2017fair}. 
In another case of a hiring application, it was recently exposed that Amazon discovered that their AI hiring system was discriminating against female candidates, particularly for software development and technical positions. One suspected reason for this is that most recorded historical data were for male software developers \citep{Dastin:2018:Online}. In a different scenario in advertising, it was shown that Google's ad-targeting algorithm had proposed higher-paying executive jobs more for men than for women \citep{datta2015automated,Simonite:2015:Online}.

These lines of evidence and concerns about algorithmic fairness have led to growing interest in the literature on defining, evaluating and improving fairness in AI algorithms (see, for example, \cite{berk2018fairness,holstein2019improving,chouldechova2018frontiers,friedler2019comparative}). 
It is important to note, however, that the task of improving fairness of AI algorithms is not trivial since there exists an inherent trade-off between accuracy and fairness.
That is, as we pursue a higher degree of fairness, we may compromise accuracy (see, for example, \cite{kleinberg2017inherent}). 

In contrast to other recent surveys in this field \cite{chouldechova2018frontiers,friedler2019comparative}, our paper proposes a comprehensive and up-to-date overview of the field, ranging from definitions and measures of fairness to state-of-the-art fairness-enhancing mechanisms.
Our survey also attempts to cover the pros and cons of the various measures and mechanisms, and guide under which setting they should be used.
Finally, a major goal of this survey is to highlight and discuss emerging areas of research that are expected to grow in the upcoming years.
Overall, this survey provides the relevant knowledge to enable new researchers to enter the field, inform current researchers on rapidly evolving sub-fields, and provide practitioners the necessary tools to apply the results.

The rest of this paper is structured as follows:
Section \ref{S:causes} discusses the potential causes of algorithmic unfairness; Section \ref{S:measures} presents definitions and measures of fairness and their trade-offs; Section \ref{S:mechanisms} reviews fairness mechanisms and methods and a comparison of the mechanisms, focusing on the pros and cons of each mechanism; Section \ref{S:datasets} outlines commonly used fairness-related datasets; Section \ref{S:emerging} presents several emerging research sub-fields of algorithmic fairness; and Section \ref{S:discussion} provides concluding remarks and sketches several open challenges for future research.

\section{Potential Causes of Unfairness}
\label{S:causes}

The literature has indicated several causes that may lead to unfairness in machine learning \citep{chouldechova2018frontiers,martinez2019fairness}:

\begin{itemize}[topsep=3pt,leftmargin=*,partopsep=3pt]

\item Biases already included in the datasets used for learning, which are based on biased device measurements, historically biased human decisions, erroneous reports or other reasons. Machine learning algorithms are essentially designed to replicate these biases.

\item Biases caused by missing data, such as missing values or sample/selection biases, which result in datasets that are not representative of the target population.

\item Biases that stem from algorithmic objectives, which aim at minimizing overall aggregated prediction errors and therefore benefit majority groups over minorities.

\item Biases caused by "proxy" attributes for sensitive attributes.
Sensitive attributes differentiate privileged and unprivileged groups, such as race, gender and age, and are typically not legitimate for use in decision making.
Proxy attributes are non-sensitive attributes that can be exploited to derive sensitive attributes.
In the case that the dataset contains proxy attributes, the machine learning algorithm can implicitly make decisions based on the sensitive attributes under the cover of using  presumably legitimate attributes \citep{barocas2016big}.

\end{itemize}

\begin{figure}[H]
\centering
\includegraphics[width=0.7\textwidth]{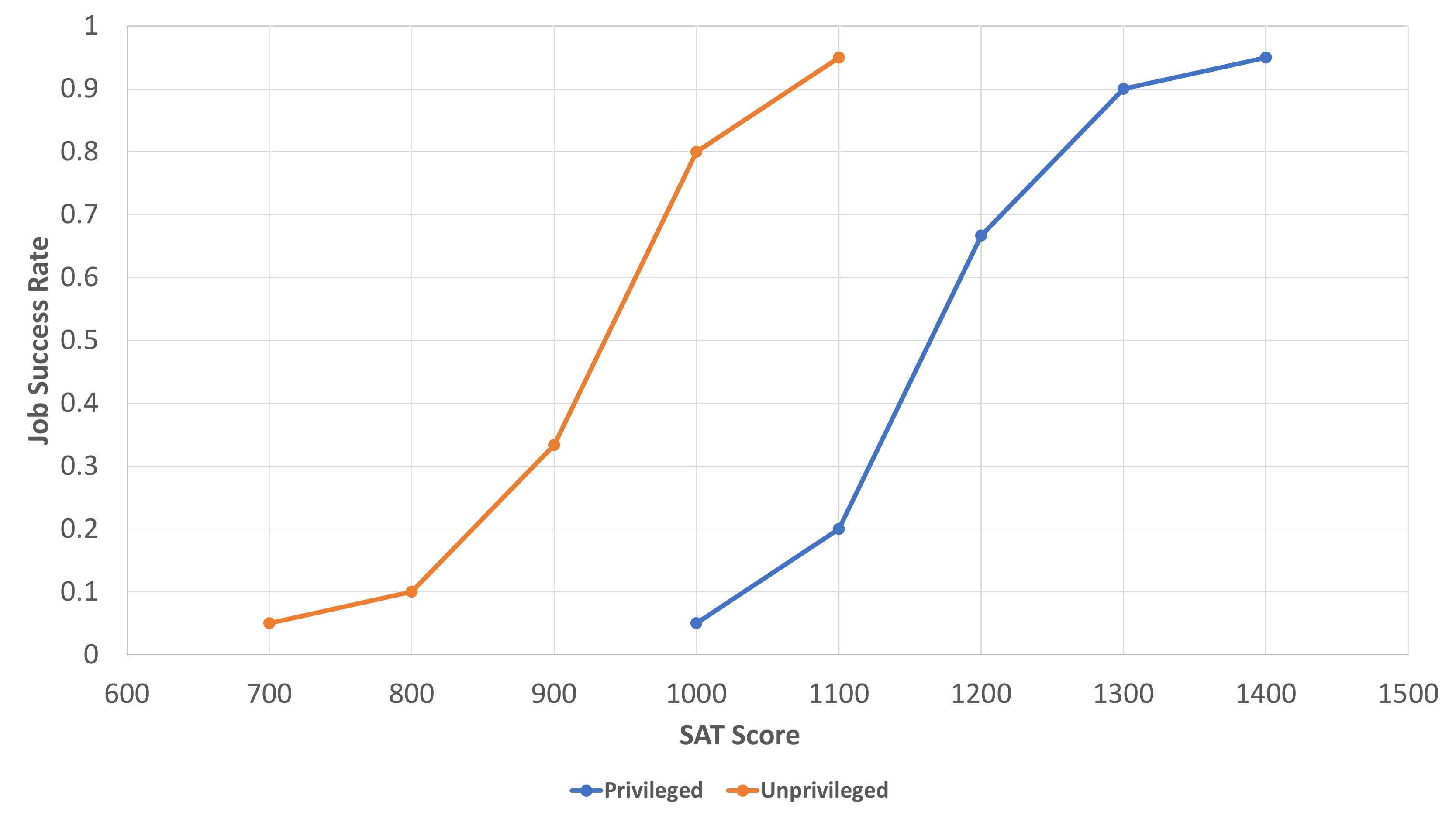}%
\caption{If the SAT scores were used for hiring, then unprivileged candidates with high potential would be excluded, whereas lower potential candidates from the privileged group would be hired instead}
\label{fig:ExampleSAT}
\end{figure}

To illustrate the last cause mentioned above, consider the example depicted in Figure \ref{fig:ExampleSAT}.
The figure illustrates a case of SAT scores for two sub-populations: a privileged one and an unprivileged one.

In this illustration, SAT scores may be used to predict the probability of job success when hiring candidates since the higher the SAT score is, the higher the probability of success. 
However, unprivileged candidates with SAT scores of approximately 1100 perform just as well as privileged candidates with SAT scores of 1400 since they may have encountered more challenging pathways to achieve their scores.
In other words, if the SAT scores were used for hiring, unprivileged candidates with high potential would be excluded, whereas lower potential candidates from the privileged group would be hired instead.

\section{Fairness Definitions and Measures}
\label{S:measures}

This section presents some general legal notions for discrimination followed by a survey of the most common measures for algorithmic fairness, and the inevitable trade-offs between them.

\subsection{Definitions of Discrimination in Legal Domains}

The legal domain has introduced two main definitions of discrimination:
i) \textbf{disparate treatment} \citep{zimmer1995emerging,barocas2016big}: intentionally treating an individual differently based on his/her membership in a protected class (\textit{direct discrimination}); ii) \textbf{disparate impact}  \citep{rutherglen1987disparate,barocas2016big}: negatively affecting members of a protected class more than others even if by a seemingly neutral policy (\textit{indirect discrimination}).

Put in our context, it is important to note that algorithms trained with data that do not include sensitive attributes (i.e., attributes that explicitly identify the protected and unprotected groups) are unlikely to produce \textit{disparate treatment}, but may still induce unintentional discrimination in the form of \textit{disparate impact} \citep{kleinberg2017inherent}.

\subsection{Measures of Algorithmic Bias}

This section presents the most prominent measures of algorithmic fairness in machine learning classification tasks.
We refer the readers to the Appendix and specifically to Table \ref{tab:measures_appendix} for a review of additional, less popular measures used in the literature.

\begin{enumerate}
\item\textbf{Disparate impact} \citep{feldman2015certifying} --
This measure was designed to mathematically represent the legal notion of \textit{disparate impact}. It requires a high ratio between the positive prediction rates of both groups. 
This ensures that the proportion of the positive predictions is similar across groups.
For example, if a positive prediction represents acceptance for a job, the condition requires the proportion of accepted applicants to be similar across groups.
Formally, this measure is computed as follows:

\begin{equation} 
\label{eq2.1}
\frac{P[\hat{Y}=1|S\neq1]}{P[\hat{Y}=1|S=1]}\geq 1-\varepsilon
\end{equation}

where $S$ represents the protected attribute (e.g., race or gender), $S=1$ is the privileged group, and $S\neq1$ is the unprivileged group. $\hat{Y}=1$ means that the prediction is positive. Let us note that if $\hat{Y}=1$ represents acceptance (e.g., for a job), then the condition requires the acceptance rates to be similar across groups. A higher value of this measure represents more similar rates across groups and therefore more fairness. Note that this notion relates to the "80 percent rule" in disparate impact law \citep{feldman2015certifying}, which requires that the acceptance rate for any race, sex, or ethnic group be at least 80\% of the rate for the group with the highest rate.

\hfill

\item\textbf{Demographic parity} -- 
This measure is similar to \textit{disparate impact}, but the difference is taken instead of the ratio \citep{calders2010three,dwork2012fairness}. This measure is also commonly referred to as \textit{statistical parity}. Formally, this measure is computed as follows:

\begin{equation} 
\label{eq2.2}
\left| P[\hat{Y}=1|S=1]-P[\hat{Y}=1|S\neq1] \right| \leq \varepsilon
\end{equation}

A lower value of this measure indicates more similar acceptance rates and therefore better fairness. \textit{Demographic parity} (and \textit{disparate impact}) ensure that the positive prediction is assigned to the two groups at a similar rate. 

\bigskip
\noindent
One disadvantage of these two measures is that a fully accurate classifier may be considered unfair, when the base rates (i.e., the proportion of actual positive outcomes) of the various groups are significantly different.
Moreover, in order to satisfy \textit{demographic parity}, two similar individuals may be treated differently since they belong to two different groups -- such treatment is prohibited by law in some cases (note that this notion also corresponds to the practice of \textit{affirmative action} \cite{sep-affirmative-action}).

\hfill

\item\textbf{Equalized odds} -- This measure was designed by \cite{hardt2016equality} to overcome the disadvantages of measures such as \textit{disparate impact} and \textit{demographic parity}. The measure computes the difference between the false positive rates (FPR), and the difference between the true positive rates (TPR) of the two groups.
Formally, this measure is computed as follows:

\begin{equation} 
\left| P[\hat{Y}=1|S=1,Y=0]-P[\hat{Y}=1|S\neq1,Y=0] \right| \leq \varepsilon
\end{equation}

\begin{equation} 
\left| P[\hat{Y}=1|S=1,Y=1]-P[\hat{Y}=1|S\neq1,Y=1] \right| \leq \varepsilon 
\end{equation}

\noindent Where the upper formula requires the absolute difference in the FPR of the two groups to be bounded by $\varepsilon$, and the lower formula requires the absolute difference in the TPR of the two groups to be bounded $\varepsilon$.
Smaller differences between groups indicate better fairness.
In contrast to \textit{demographic parity} and \textit{disparate impact} measures, a fully accurate classifier will necessarily satisfy the two \textit{equalized odds} constraints.
Nevertheless, since \textit{equalized odds} relies on the actual ground truth (i.e., $Y$), it assumes that the base rates of the two groups are representative and were not obtained in a biased manner.

\hfill

One use case that demonstrates the effectiveness of this measure investigated the COMPAS \citep{compas} algorithm used in the United States criminal justice system. For predicting recidivism, although its accuracy was similar for both groups (African-Americans and Caucasians), it was discovered that the \textit{odds} were different. It was discovered that the system had falsely predicted future criminality (FPR) among African-Americans at twice the rate predicted for white people \citep{Angwin:2016:Online}; importantly, the algorithm also induced the opposite error, significantly underestimating future crimes among Caucasians (FNR).

\hfill

\item\textbf{Equal opportunity} -- 
This requires true positive rates (TPRs) to be similar across groups (meaning the probability of an individual with a positive outcome to have a positive prediction) \citep{hardt2016equality}. This measure is similar to equalized odds but focuses on the true positive rates only. This measure is mathematically formulated as follows:

\begin{equation} 
\label{eq2.4}
\left| P[\hat{Y}=1|S\neq1,Y=1]-P[\hat{Y}=1|S=1,Y=1]\right| \leq \varepsilon
\end{equation}

Let us note that following the equality in terms of only one type of error (e.g., true positives) will increase the disparity in terms of the other error \citep{pleiss2017fairness}.
Moreover, according to \cite{corbett2018measure}, this measure may be problematic when base rates differ between groups.

\hfill

\newcounter{enumTemp}
\setcounter{enumTemp}{\theenumi}
\end{enumerate}

\noindent Thus far, we have mapped the most common \textit{group} notions of fairness, which require parity of some statistical measure across groups.
The literature has additionally indicated \textit{individual} notions of fairness.
It is alternatively possible to match other measures such as accuracy, error rates or calibration values between groups (see the Appendix and specifically Table \ref{tab:measures_appendix}).
\textit{Group} definitions of fairness, such as \textit{demographic parity, disparate impact, equalized odds} and \textit{equalized opportunity}, consider fairness with respect to the whole group, as opposed to \textit{individual} notions of fairness.

\hfill
\begin{enumerate}
\setcounter{enumi}{\theenumTemp}

\item\textbf{Individual fairness} -- This requires that similar individuals will be treated similarly. Similarity may be defined with respect to a particular task \citep{dwork2012fairness,joseph2016fairness}.
Individual fairness may be described as follows:

\begin{equation} 
\label{eq2.5}
\left| P(\hat{Y}^{(i)} = y | X^{(i)},S^{(i)}) - P(\hat{Y}^{(j)} = y|{X}^{(j)},S^{(j)}) \right| \leq  \varepsilon; \;
if \; d(i, j) \approx 0
\end{equation}

where i and j denote two individuals, $S^{(\cdot)}$ refers to the individuals' sensitive attributes and $X^{(\cdot)}$ refers to their associated features.
$d(i, j)$ is a distance metric between individuals that can be defined depending on the domain such that similarity is measured according to an intended task. This measure considers other individual attributes for defining fairness, rather than just the sensitive attributes.
However, note that in order to define similarity between individuals, a similarity metric needs to be defined, which is not trivial. This measure, in addition to assuming a similarity metric, also requires some assumptions regarding the relationship between features and labels (see, for example, \cite{chouldechova2018frontiers}).

\end{enumerate}

\subsection{Trade-offs}

Determining the right measure to be used must take into account the proper legal, ethical, and social context.
As demonstrated above, different measures exhibit different advantages and disadvantages.
Next, we highlight the main trade-offs that exist between different notions of fairness, and the inherent trade-off between fairness and accuracy.

\subsection*{Fairness measures trade-offs}

Interestingly, several recent studies have shown that it is not possible to satisfy multiple notions of fairness simultaneously \citep{berk2018fairness,chouldechova2017fair,friedler2016possibility,kleinberg2017inherent,pleiss2017fairness,corbett2017algorithmic,corbett2018measure}. For example, when base rates differ between groups, it is not possible to have a classifier that equalizes both calibration and odds (except for trivial cases such as a classifier that assigns all examples to a single class). Additionally, there is also evidence for incompatibility between equalized accuracy and equalized odds, as in the COMPAS criminal justice use case \citep{Angwin:2016:Online, berk2018fairness}.

\cite{pleiss2017fairness} recommends that in light of the inherent incompatibility between equalized calibration and equalized odds, practical implications requires choosing only one of these goals according to the specific application's requirements. We recommend that any selected measure of algorithmic fairness be considered in the appropriate legal, social and ethical contexts.

\begin{table}[H]
\fontsize{7}{12}\selectfont
\centering
\renewcommand{\arraystretch}{0.9}
\renewcommand\baselinestretch{0.65}\selectfont
\centering
\caption{Measures and Definitions for Algorithmic Fairness}
\label{tab:measures}
\begin{tabular}
{|p{0.095\columnwidth}||p{0.03\columnwidth}|p{0.15\columnwidth}|p{0.065\columnwidth}|p{0.07\columnwidth}|p{0.08\columnwidth}|p{0.08\columnwidth}|p{0.08\columnwidth}|p{0.18\columnwidth}|} 
\hline
\multicolumn{1}{|l||}{\textbf{Measure}} 
& \multicolumn{1}{l|}{\textbf{Paper}} 
& \multicolumn{1}{l|}{\textbf{Description}} 
& \multicolumn{1}{l|}{\textbf{Type}} 
& \textbf{Uses Actual Outcome} 
& \textbf{Uses Sensitive Attribute} 
& \textbf{Type of Actual Outcome} 
& \textbf{Type of Sensitive Attribute} 
& \multicolumn{1}{l|}{\textbf{Equivalent Notions}} 
\\ \hline \hline
\textbf{Disparate Impact} & \cite{ feldman2015certifying} 
& High ratio between positive prediction rates of both groups & Group 
& \xmark & \cmark & - & Binary
& For $\varepsilon=0.2$ relates to the "80 percent rule" in disparate impact law \citep{feldman2015certifying} 
\\ \hline
\textbf{Demographic Parity} & \cite{calders2010three}, \cite{dwork2012fairness} 
& Similar positive prediction rates between groups & Group 
& \xmark & \cmark & - & Binary
& \vspace{-4pt}\begin{itemize}[noitemsep,topsep=0pt,leftmargin=*,partopsep=0pt]
\item Statistical parity \citep{dwork2012fairness};
\item Group fairness \citep{dwork2012fairness};
\item Equal acceptance rates \citep{vzliobaite2017measuring,verma2018fairness,chouldechova2017fair};
\item Discrimination score \cite{calders2010three} 
\vspace{-\baselineskip}\end{itemize}
\\ \hline
\textbf{Equal Opportunity} & \cite{hardt2016equality} 
& Requires that TPRs are similar across groups 
& Group 
& \cmark & \cmark & Binary & Binary
& \vspace{-4pt}\begin{itemize}[noitemsep,topsep=0pt,leftmargin=*,partopsep=0pt]
\item Equal true positive rate (TPR);
\item Mathematically equal TPRs will induce equal false negative rates (FNRs) (see \cite{verma2018fairness,corbett2017algorithmic});
\item False negative error rate balance \citep{chouldechova2017fair,verma2018fairness};
\vspace{-\baselineskip}\end{itemize}
\\ \hline
\textbf{Equalized Odds} & \cite{hardt2016equality} 
& Requires that FPRs (1-TNR) and TPRs (1-FNR) are similar across groups & Group 
& \cmark & \cmark & Binary & Binary
& \vspace{-4pt}\begin{itemize}[noitemsep,topsep=0pt,leftmargin=*,partopsep=0pt]
\item Disparate mistreatment \citep{zafar2017fairness};
\item Error rate balance \citep{chouldechova2017fair};
\item Conditional procedure accuracy equality \citep{berk2018fairness}
\vspace{-\baselineskip}\end{itemize}
\\ \hline
\textbf{Fairness through Awareness} & \cite{dwork2012fairness} 
& Requires that similar individuals will have similar classifications. Similarity can be defined with respect to a specific task
& Individual 
& \xmark & \xmark & - & -
& Individual fairness 
\\ \hline
\end{tabular}
\end{table}
\restoregeometry

\subsection*{Fairness-accuracy trade-off}

The literature extensively discusses the inherent trade-off between accuracy and fairness - as we pursue a higher degree of fairness, we may compromise accuracy (see for example \cite{kleinberg2017inherent}). 
A theoretical analysis of the trade-off between fairness and accuracy was studies in \cite{corbett2017algorithmic} and \cite{lipton2017does}. Since then, many papers have empirically supported the existence of this trade-off (for example, \cite{friedler2019comparative,menon2018cost,bechavod2017learning}).
Generally, the aspiration of a fairness-aware algorithm is to achieve a model that allows for higher fairness without significantly compromising the accuracy or other alternative notions of utility.

\bigskip
\noindent
Table \ref{tab:measures} presents a summary of the measures presented in this section.
For further reading about algorithmic fairness measures, we refer the reader to \cite{corbett2018measure}, \cite{kleinberg2017inherent}, and \cite{verma2018fairness}.

\section{Fairness-Enhancing Mechanisms}
\label{S:mechanisms}

Numerous recent papers have proposed mechanisms to enhance fairness in machine learning algorithms.
These mechanisms are typically categorized into three types: pre-process, in-process, and post-process.
The following three subsections review studies in each one of these categories.
The fourth subsection is devoted for comparing the three mechanism types and providing guidelines on when each type should be used.

\subsection{Pre-Process Mechanisms}

Mechanisms in this category involve changing the training data before feeding it into a machine learning algorithm. Preliminary mechanisms, such as the ones proposed by \cite{kamiran2012data} and \cite{luong2011k} proposed changing the labels of some instances or reweighing them before training to make the classification fairer. Typically, the labels that are changed are related to samples that are closer to the decision boundary since these are the ones that are most likely to be discriminated. More recent mechanisms suggest modifying feature representations, so that a subsequent classifier will be fairer \citep{louizos2016variational,feldman2015certifying,calmon2017optimized,zemel2013learning,samadi2018price}.

For example, \cite{feldman2015certifying} suggest modifying the features in the dataset so that the distributions for both privileged and unprivileged groups become similar, and therefore, making it more difficult for the algorithm to differentiate between the two groups.
A tuning parameter $\lambda$ was provided for controlling the trade-off between fairness and accuracy ($\lambda$=0 indicates no fairness considerations, while $\lambda$=1 maximizes fairness).
\cite{chierichetti2017fair,backurs2019scalable} use the same notion of fair representation learning and applies it for fair clustering, and \cite{samadi2018price} applies it for fair dimensionality reduction (PCA). For more fair representation learning using \textit{adversarial learning}, see section \ref{S:emerging_adversarial}.

Note that this approach to achieving fairness is somewhat related to the field of data \textit{compression} \citep{tishby1999information,zemel2013learning}. It is also very closely related to \textit{privacy} research since both fairness and privacy can be enhanced by removing or obfuscating the sensitive information, with the adversary goal of minimal data distortion \citep{edizel2019fairecsys,kazemi2018scalable}.

\subsection{In-Process Mechanisms}

These mechanisms involve modifying the machine learning algorithms to account for fairness during the training time \citep{kamishima2012fairness,woodworth2017learning,bechavod2017learning,bechavod2017penalizing,zafar2017AIAS,zafar2017fairness,goh2016satisfying,calders2010three,agarwal2018reductions}.

For example, \cite{kamishima2012fairness} suggest adding a regularization term to the objective function that penalizes the mutual information between the sensitive feature and the classifier predictions. A tuning parameter $\eta$ was provided to modulate the trade-off between fairness and accuracy.

\cite{zafar2017fairness}, \cite{zafar2017AIAS} and \cite{woodworth2017learning} suggest adding constraints to the classification model that require satisfying a proxy for \textit{equalized odds} \citep{zafar2017fairness,woodworth2017learning} or \textit{disparate impact} \citep{zafar2017AIAS}. \cite{woodworth2017learning} also show that there exist difficult computational challenges in learning a fair classifier based on \textit{equalized odds}.

\cite{bechavod2017learning} and \cite{bechavod2017penalizing} suggest incorporating penalty terms into the objective function that enforce matching proxies of FPR and FNR. 
\cite{kamiran2010discrimination} suggest adjusting a decision tree split criterion to maximize information gain between the split attribute and the class label while minimizing information gain with respect to the sensitive attribute. \cite{zemel2013learning} combine fair representation learning with an in-process model by applying a multi-objective loss function based on logistic regression, and \cite{louizos2016variational} apply this notion using a variational autoencoder.

\cite{quadrianto2017recycling} suggest using the notion of \textit{privileged learning}\footnotemark{} for improving performance in cases where the sensitive information is available at training time but not at testing time. They add constraints and regularization components to the privileged learning support vector machine (SVM) model proposed by \cite{vapnik2015learning}. They combine the sensitive attributes as privileged information that is known only at training time, and they additionally use a maximum mean discrepancy (MMD) criterion \citep{gretton2012kernel} to encourage the distributions to be similar across privileged and unprivileged groups.

\footnotetext[1]{Privileged learning is designed to improve performance by using additional information, denoted as the “privileged information,” which is present only in the training stage and not in the testing stage \citep{vapnik2015learning}.}

\cite{berk2017convex} propose a convex in-process fairness mechanism for regression tasks and use three regularization terms that include variations of individual fairness, group fairness and a combined hybrid fairness penalty term. \cite{agarwal2019fair} propose an in-process minimax optimization formulation for enhancing fairness in regression tasks based on the suggested design of \cite{agarwal2018reductions} for classification tasks. They use two fairness metrics adjusted for regression tasks. One is an adjusted \textit{demographic parity} measure, which requires the predictor to be independent of the sensitive attribute as measured by the cumulative distribution function (CDF) of the protected group compared to the CDF of the general population \citep{agarwal2019fair} using the \textit{Kolmogorov-Smirnov statistic} \citep{lehmann2006testing}. The second measure is the bounded group loss (BGL), which requires that the prediction error of all groups remain below a predefined level \citep{agarwal2019fair}.

\subsection{Post-Process Mechanisms}

These mechanisms perform post-processing of the output scores of the classifier to make decisions fairer \citep{corbett2017algorithmic, dwork2018decoupled,hardt2016equality,menon2018cost}.
For example, \cite{hardt2016equality} propose a technique for flipping some decisions of a classifier to enhance equalized odds or equalized opportunity. \cite{corbett2017algorithmic} and \cite{menon2018cost} similarly suggest selecting separate thresholds for each group separately, in a manner that maximizes accuracy and minimizes demographic parity.
\cite{dwork2018decoupled} propose a decoupling technique to learn a different classifier for each group.
They additionally combine a \textit{transfer learning} technique with their procedure to learn from out-of-group samples (to read more about transfer learning, see \cite{pan2009survey}).

\bigskip
\noindent
Table \ref{tab:methods} presents a summary of the pre-process, in-process and post-process mechanisms for algorithmic fairness discussed in this section.
These methods were designed for the task of classification. 
Fairness mechanisms for other learning tasks are discussed in section \ref{S:emerging}.

\begin{table}[H]
\fontsize{7}{12}\selectfont
\centering
\renewcommand{\arraystretch}{0.8}
\renewcommand\baselinestretch{0.65}\selectfont
\caption{Pre-Process, In-Process and Post-Process Mechanisms for Algorithmic Fairness}
\label{tab:methods}
\begin{tabular}
{|p{0.05\columnwidth}||p{0.08\columnwidth}|p{0.08\columnwidth}|p{0.17\columnwidth}|p{0.17\columnwidth}|p{0.17\columnwidth}|p{0.14\columnwidth}|} \hline 
\hline
\textbf{Paper} 
& \textbf{Mechanism Type} 
& \textbf{Base Algorithm} 
& \textbf{Optimization Measure}
& \textbf{Evaluation Measure}
& \textbf{Method Name}
& \textbf{Datasets}
\\ \hline \hline
\cite{kamishima2012fairness} 
& In-Process 
& Logistic regression 
& Mutual information between prediction and sensitive attribute
& Normalized prejudice index
& Prejudice Remover Regularizer
& Adult (test only) 
\\ \hline
\cite{feldman2015certifying} 
& Pre-Process 
& Any 
& Earth moving distance
& Disparate impact
& Removing Disparate Impact
& 
\vspace{-4pt}\begin{itemize}[noitemsep,topsep=0pt,leftmargin=*,partopsep=0pt]
\item Adult
\item German
\vspace{-4pt}\end{itemize}
\\ \hline
\cite{zafar2017AIAS} 
& In-Process 
& Decision boundary-based 
& Covariance (between sensitive attributes and distance to the decision boundary)
& Disparate impact
& Fairness Constraints
& ProPublica
\\ \hline
\cite{zafar2017fairness} 
& In-Process 
& Decision boundary-based 
& Proxy for equalized odds
& Equalized odds
& Removing Disparate Mistreatment
& ProPublica
\\ \hline
\cite{bechavod2017penalizing}
& In-Process 
& Decision boundary-based 
& Proxy for equalized odds
& Equalized odds
& Penalizing Unfairness
& 
\vspace{-4pt}\begin{itemize}[noitemsep,topsep=0pt,leftmargin=*,partopsep=0pt]
\item ProPublica
\item Adult
\item Loans
\item Admissions
\vspace{-4pt}\end{itemize}
\\ \hline
\cite{kamiran2010discrimination} 
& In-Process, Post-Process 
& In-Process - decision tree; Post-Process - any algorithm
& Information gain
& Demographic parity
& \vspace{-4pt}
\begin{itemize}[noitemsep,topsep=0pt,leftmargin=*,partopsep=0pt] \item Discrimination Aware Tree Construction (new split criterion); \item Relabeling (post-process)
\vspace{-\baselineskip}\end{itemize}
& 
\vspace{-4pt}\begin{itemize}[noitemsep,topsep=0pt,leftmargin=*,partopsep=0pt]
\item Adult
\item Communities
\item Dutch census 
\vspace{-4pt}\end{itemize}
\\ \hline 
\cite{kamiran2012data} 
& Pre-Process 
& Any score-based 
& Acceptance probabilities, distance from boundary
& Demographic parity
& \vspace{-4pt}
\begin{itemize}[noitemsep,topsep=0pt,leftmargin=*,partopsep=0pt]
\item Massaging
\item Reweighing
\item Sampling
\item Suppression
\vspace{-\baselineskip}\end{itemize}
& 
\vspace{-4pt}\begin{itemize}[noitemsep,topsep=0pt,leftmargin=*,partopsep=0pt]
\item German
\item Adult
\item Communities
\item Dutch
\vspace{-4pt}\end{itemize}
\\ \hline 
\cite{calders2010three} 
& In-Process, Post-Process 
& Naive Bayes 
& Acceptance probabilities
& Demographic parity
& \vspace{-4pt}\begin{itemize}[noitemsep,topsep=0pt,leftmargin=*,partopsep=0pt]
\item Modifying Naive Bayes
\item Two Naive Bayes
\item Expectation Maximization
\vspace{-4pt}\end{itemize}
& Adult (test only) 
\\ \hline
\cite{luong2011k} 
& Pre-Process 
& Any 
& Conditional statistical parity
& Conditional statistical parity
& Discrimination Prevention with KNN
& 
\vspace{-4pt}\begin{itemize}[noitemsep,topsep=0pt,leftmargin=*,partopsep=0pt]
\item Adult
\item Communities
\item German
\vspace{-4pt}\end{itemize}
\\ \hline \cite{zemel2013learning} 
& Pre-Process + In-Process 
& Logistic regression 
& Demographic parity
& Demographic parity
& Learning Fair Representations
& 
\vspace{-4pt}\begin{itemize}[noitemsep,topsep=0pt,leftmargin=*,partopsep=0pt]
\item Adult
\item German
\item Heritage health
\vspace{-4pt}\end{itemize}
\\ \hline
\cite{goh2016satisfying} 
& In-Process 
& SVM 
& Disparate impact, Equal opportunity
& Disparate impact
& Dataset Constraints
& Adult
\\ \hline
\cite{hardt2016equality} 
& Post-Process 
& Any score-based 
& Equalized odds
& Equalized odds
& Equality of Opportunity in Supervised Learning
& FICO scores \citep{hardt2016equality} 
\\ \hline
\cite{corbett2017algorithmic} 
& Post-Process 
& Any score-based 
& \vspace{-4pt}\begin{itemize}[noitemsep,topsep=0pt,leftmargin=*,partopsep=0pt]
\item Demographic parity
\item Conditional statistical parity
\item Predictive parity
\vspace{-4pt}\end{itemize}
& \vspace{-4pt}\begin{itemize}[noitemsep,topsep=0pt,leftmargin=*,partopsep=0pt]
\item Demographic parity
\item Conditional statistical parity
\item Predictive parity
\vspace{-4pt}\end{itemize}
& Cost of Fairness
& ProPublica 
\\ \hline
\cite{woodworth2017learning} 
& In-Process + Post-Process 
& Convex linear 
& Equalized odds
& Equalized odds
& Learning Non-Discriminatory Predictors
& - 
\\ \hline
\cite{quadrianto2017recycling} 
& In-Process 
& SVM 
& 
Maximum mean discrepancy (MMD)
& 
\vspace{-4pt}\begin{itemize}[noitemsep,topsep=0pt,leftmargin=*,partopsep=0pt]
\item Equalized odds
\item Overall accuracy equality 
\vspace{-4pt}\end{itemize}
& Recycling Privileged Learning and Distribution Matching
& 
\vspace{-4pt}\begin{itemize}[noitemsep,topsep=0pt,leftmargin=*,partopsep=0pt]
\item ProPublica
\item Adult
\vspace{-4pt}\end{itemize}
\\ \hline
\cite{calmon2017optimized} 
& Pre-Process 
& Any 
& 
\vspace{-4pt}\begin{itemize}[noitemsep,topsep=0pt,leftmargin=*,partopsep=0pt]
\item Disparate impact
\item Individual fairness
\vspace{-4pt}\end{itemize}
& Disparate impact
& Optimized Pre-processing for Discrimination Prevention
& 
\vspace{-4pt}\begin{itemize}[noitemsep,topsep=0pt,leftmargin=*,partopsep=0pt]
\item ProPublica
\item Adult
\vspace{-4pt}\end{itemize}
\\ \hline
\cite{dwork2018decoupled} 
& In-Process + Post-Process 
& Any score-based 
& \vspace{-4pt}\begin{itemize}[noitemsep,topsep=0pt,leftmargin=*,partopsep=0pt]
\item Demographic parity
\item Equalized odds
\vspace{-4pt}\end{itemize}
& \vspace{-4pt}\begin{itemize}[noitemsep,topsep=0pt,leftmargin=*,partopsep=0pt]
\item Demographic parity
\item Equalized odds
\vspace{-4pt}\end{itemize}
& Decoupled Classifiers
& ImageNet \citep{deng2009imagenet} 
\\ \hline
\cite{menon2018cost} 
& Post-Process 
& Any score-based 
& \vspace{-4pt}\begin{itemize}[noitemsep,topsep=0pt,leftmargin=*,partopsep=0pt]
\item Demographic parity
\item Equal opportunity
\vspace{-4pt}\end{itemize}
& \vspace{-4pt}\begin{itemize}[noitemsep,topsep=0pt,leftmargin=*,partopsep=0pt]
\item Demographic parity
\item Equal opportunity
\vspace{-4pt}\end{itemize}
& 
Plugin Approach
& - 
\\ \hline
\cite{agarwal2018reductions} 
& In-Process 
& Any 
& \vspace{-4pt}\begin{itemize}[noitemsep,topsep=0pt,leftmargin=*,partopsep=0pt]
\item Demographic parity
\item Equalized odds
\vspace{-4pt}\end{itemize}
& \vspace{-4pt}\begin{itemize}[noitemsep,topsep=0pt,leftmargin=*,partopsep=0pt]
\item Demographic parity
\item Equalized odds
\vspace{-4pt}\end{itemize}
& Reductions Approach
& 
\vspace{-4pt}\begin{itemize}[noitemsep,topsep=0pt,leftmargin=*,partopsep=0pt]
\item ProPublica
\item Adult
\item Dutch
\item Admissions
\vspace{-4pt}\end{itemize}
\\ \hline
\cite{louizos2016variational} 
& Pre-Process + In-Process 
& Any 
& 
Maximum mean discrepancy (MMD)
& \vspace{-4pt}\begin{itemize}[noitemsep,topsep=0pt,leftmargin=*,partopsep=0pt]
\item Demographic parity
\item Mean difference
\vspace{-4pt}\end{itemize}
& Variational Fair Autoencoder
& 
\vspace{-4pt}\begin{itemize}[noitemsep,topsep=0pt,leftmargin=*,partopsep=0pt]
\item Adult
\item German
\item Heritage health
\vspace{-4pt}\end{itemize}
\\ \hline
\end{tabular}
\end{table}

\subsection{Which Mechanism to Use?}
\label{S:benchmark}

The different mechanism types present respective advantages and disadvantages.
Pre-process mechanisms can be advantageous since they can be used with any classification algorithm.
However, they may harm the explainability of the results.
Moreover, since they are not tailored for a specific classification algorithm, there is high uncertainty with regard to the level of accuracy obtained at the end of the process.

Similar to pre-process mechanisms, post-process mechanisms may be used with any classification algorithm. However, due to the relatively late stage in the learning process in which they are applied, post-process mechanisms typically obtain inferior results \citep{woodworth2017learning}.
In a post-process mechanism, it may be easier to fully remove bias types such as \textit{disparate impact}; however, this is not always the desired measure, and it could be considered as discriminatory since it deliberately damages accuracy for some individuals in order to compensate others (this is also related to the controversies in the legal and economical field of \textit{affirmative action}, see \cite{sep-affirmative-action}).
Specifically, post-process mechanisms may treat differently two individuals who are similar across all features except for the group to which they belong. This approach requires the decision maker at the end of the loop to possess the information of the group to which individuals belong (this information may be unavailable due to legal or privacy reasons).

In-process mechanisms are beneficial since they can explicitly impose the required trade-off between accuracy and fairness in the objective function \citep{woodworth2017learning}. However, such mechanisms are tightly coupled with the machine algorithm itself.

Hence, we see that the selection of method depends on the availability of the ground truth, the availability of the sensitive attributes at test time, and on the desired definition of fairness, which can also vary from one application to another.

\hfill

\noindent Several preliminary attempts were made in order to understand which methods are best for use. The study in \cite{hamilton2017benchmarking} was a first effort in comparing several fairness mechanisms previously proposed in the literature \citep{kamishima2012fairness,feldman2015certifying,calders2010three,zafar2017AIAS}. The analysis focuses on binary classification with binary sensitive attributes. The authors have demonstrated that the performances of the methods vary across datasets, and there was no conclusively dominating method.

Another study by \cite{roth2018comparison} has shown as a preliminary benchmark that in several cases, in-process mechanisms perform better than pre-process mechanisms, and for other cases, they do not, leading to the conclusion that there is a need for much more extensive experiments.

A recent empirical study \citep{friedler2019comparative} has provided a benchmark analysis of several fairness-aware methods and compared the fairness-accuracy trade-offs obtained by these methods. The authors have tested the performances of these methods across different measures of fairness and across different datasets.
They have concluded that there was no single method that outperformed the others in all cases and that the results depend on the fairness measure, on the dataset, and on changes in the train-test splits.

More research is required for developing robust fairness mechanisms and metrics or, alternatively, for finding the adequate mechanism and metric for each scenario. For instance, the conclusions reached when considering missing data might be very different than those reached when all information is available \citep{kallus2019assessing,martinez2019fairness}.
\cite{kallus2019assessing} explore the limitations of measuring fairness when the membership in a protected group is not available in the data. \cite{martinez2019fairness} have tested imputation strategies to deal with the fairness of partially missing examples in the dataset. They have shown that rows containing missing values may be more fair than the rest and therefore suggest imputation rather than deletion of these data.
\cite{Pessach2020fairselection} find that when there is an evident selection bias in the data, meaning that there is an extreme under-representation of unprivileged groups, pre-process mechanisms can outperform in-process mechanisms.

\section{Fairness-Related Datasets}
\label{S:datasets}

In this section, we review the most commonly used datasets in the literature of algorithmic fairness.

\subsection*{ProPublica risk assessment dataset}

The ProPublica dataset includes data from the COMPAS risk assessment system (see \cite{compas,Angwin:2016:Online,Larson:2016:Online}). 

This dataset was previously extensively used for fairness analysis in the field of criminal justice risk \citep{berk2018fairness}. The dataset includes 6,167 individuals, and the features in the dataset include number of previous felonies, charge degree, age, race and gender. The target variable indicates whether an inmate recidivated (was arrested again) within two years after release from prison.

As for the sensitive variable, this dataset was previously used with two variations -- the first when race was considered as the sensitive attribute and the second when gender was considered as the sensitive attribute \citep{friedler2019comparative,martinez2019fairness,calmon2017optimized,emelianov2019price,bechavod2017penalizing}.

\subsection*{Adult income dataset}

The Adult dataset is a publicly available dataset in the UCI repository \citep{Dua:2019} based on 1994 US census data. The goal of this dataset is to successfully predict whether an individual earns more or less than 50,000\$ per year based on features such as occupation, marital status, and education.
The sensitive attributes in this dataset includes age \citep{louizos2016variational}, gender \citep{zemel2013learning} and race \citep{zafar2017AIAS,friedler2019comparative,martinez2019fairness}.

This dataset is used with several different preprocessing procedures. For example, the dataset of \cite{zafar2017AIAS} includes 45,222 individuals after preprocessing (48,842 before preprocessing).

\subsection*{German credit dataset}

The German dataset is a publicly available dataset in the UCI repository \citep{Dua:2019} that includes information of individuals from a German bank in 1994.

The goal of this dataset is to predict whether an individual should receive a good or bad credit risk score based on features such as employment, housing, savings, and age. 
The sensitive attributes in this dataset include gender \citep{louizos2016variational,friedler2019comparative} 
and age \citep{zemel2013learning,kamiran2009classifying}. 
This dataset is significantly smaller, with only 1,000 individuals with 20 attributes.

\subsection*{Ricci promotion dataset}

The Ricci dataset includes the results of an exam administered to 118 individuals to determine which of them would receive a promotion. The dataset originated from a case that was brought to the United States Supreme Court \citep{miao2011did,rutherglen2009ricci}.
The goal of this dataset is to successfully predict whether an individual receives a promotion based on features that were tested in the exam, as well as the current position of each individual. The sensitive attribute in this dataset is race.

\subsection*{Mexican poverty dataset}

The Mexican poverty dataset includes poverty estimation for determining whether to match households with social programs. The data originated from a survey of 70,305 households in 2016 \citep{ibarraran2017conditional}. 
The target feature is poverty level, and there are 183 features.
This dataset was studied, for example, in \cite{noriega2019active}.
The authors studied two sensitive features: young and old families; urban and rural areas.

\subsection*{Diabetes dataset}

The Diabetes dataset includes hospital data for the task of predicting whether a patient will be readmitted. It is publicly available in the UCI repository \citep{Dua:2019}. The data contain approximately 100,000 instances and 235 attributes.
This dataset was studied, for example, in \cite{edwards2015censoring}, where it was studied with race as the sensitive feature.

\subsection*{Heritage health dataset}

The Heritage health dataset originated from a competition conducted by the United States as a competition to improve healthcare through early prediction.
It includes data of 147,473 patients with 139 features. The goal of this dataset is to predict whether an individual will spend any days in the hospital during the next year \citep{brierley2011heritage}. This dataset was studied, for example, in \cite{zemel2013learning}, \cite{louizos2016variational} and \cite{tramer2017fairtest}, where age was the sensitive feature.

\subsection*{The College Admissions dataset}

The College Admissions dataset was collected by the UCLA law school \citep{sander2004systemic}.
It includes data of over 20,000 records of law school students who took the bar exam.
The goal of this dataset is to predict whether a student will pass the exam based on factors such as LSAT score, undergraduate GPA and family income.

This dataset was used, for example, by \cite{berk2017convex}, where gender was studied as the sensitive feature, and \cite{bechavod2017penalizing}, where race was studied as the sensitive feature.

\subsection*{The Bank Marketing dataset}

\noindent The Bank Marketing dataset is a publicly available dataset in the UCI repository \citep{Dua:2019,moro2014data}, and it includes 41,188 individuals with 20 attributes. The task is to predict whether the client has subscribed to a term deposit service based on features such as marital status and age.
It was previously investigated by \cite{zafar2017AIAS}, where age was studied as the sensitive attribute.

\subsection*{The Loans Default dataset}

The Loans Default dataset includes 30,000 instances and 24 attributes of credit card users. It is publicly available in the UCI repository \citep{Dua:2019,yeh2009comparisons}.
The goal is to predict whether a customer will default on payments. The features include age, gender, marital status, past payments, credit limit and education.

This dataset was used, for example, by \cite{bechavod2017penalizing} and \cite{yeh2009comparisons}, where gender was studied as the sensitive feature.

\subsection*{The Dutch Census dataset}

The Dutch Census dataset includes 189,725 instances and 13 attributes of individuals. It is publicly available in the IPUMS repository \citep{minnesota2015integrated}.
\cite{kamiran2012data} and \cite{agarwal2018reductions} use this dataset with only the 60,420 individuals who are not underaged. Their goal is to predict whether an individual holds a highly prestigious occupation by using features such as gender, age, household details, location, citizenship, birth country, education, economic status, and marital status. The sensitive feature utilized is gender.

\subsection*{The Communities and Crimes dataset}

The Communities and Crimes dataset includes 1,994 instances and 128 attributes of communities in the United States.
It is publicly available in the UCI repository \citep{redmond2002data,Dua:2019}.
The goal is to predict the number of violent crimes per 100,000 individuals based on features such as percentage of population by age, by marital status, by number of children, by race, and more.
\cite{kamiran2012data} add a new sensitive attribute that represents whether the percentage of the African-American population in the community is greater than 0.06.

\bigskip
\noindent
Table \ref{tab:datasets} presents a summary of the benchmark datasets for algorithmic fairness that were discussed in this section.

\begin{table}[H]
\centering
\caption{Common Benchmark Datasets for Algorithmic Fairness}
\label{tab:datasets}
\begin{tabular}
{|p{0.15\columnwidth}||p{0.12\columnwidth}|p{0.11\columnwidth}|p{0.25\columnwidth}|p{0.3\columnwidth}|}
\hline 
\textbf{Dataset Name} & \textbf{Domain} & \textbf{\# Records} & \textbf{Sensitive Attributes} & \multicolumn{1}{l|}{\textbf{Target Attributes}} \\ \hline \hline
\textbf{ProPublica} & Criminal risk assessment & 6,167 & Race; Gender & Whether an inmate has recidivated (was arrested again) in less than two years after release from prison \\ \hline
\textbf{Adult} & Income & 48,842 & Age; Gender & Whether an individual earns more or less than 50,000\$ per year \\ \hline
\textbf{German} & Credit & 1,000 & Gender; Age & Whether an individual should receive a good or bad credit risk score \\ \hline
\textbf{Ricci} & Promotion & 118 & Race & Whether an individual receives a promotion \\ \hline
\textbf{Mexican poverty} & Poverty & 183 & Young and old families; Urban and rural areas & Poverty level of households \\ \hline
\textbf{Diabetes} & Health & 100,000 & Race & Whether a patient will be readmitted \\ \hline
\textbf{Heritage health} & Health & 147,473 & Age & Whether an individual will spend any days in the hospital in the next year \\ \hline
\textbf{College Admissions} & College Admissions & 20,000 & Gender; Race & Whether a law student will pass the bar exam \\ \hline
\textbf{Bank Marketing} & Marketing & 41,188 & Age & Whether the client subscribed to a term deposit service \\ \hline
\textbf{Loans Default} & Loans & 30,000 & Gender & Whether a customer will default on payments \\ \hline
\textbf{Dutch Census} & Census & 189,725 & Gender & Whether an individual holds a highly prestigious occupation \\ \hline
\textbf{Communities and Crimes} & Crime & 1,994 & Percentage of African-American population & For each community, the number of violent crimes per 100,000 individuals \\ \hline
\end{tabular}
\end{table}

\section{Emerging Research on Algorithmic Fairness}
\label{S:emerging}

In this section, we review selected emerging sub-fields of algorithmic fairness.

\subsection{Fair Sequential Learning}

Most existing research on algorithmic fairness considers batch classification, where the complete data are available in advance. However, many applications require investigation of online and reinforcement learning, where the data are collected over time. In online learning, in contrast to batch learning, the system includes feedback loops so that the decision at each step may influence the state and future decisions. This imposes challenges in both defining and making fair algorithmic decisions, as fairness should now be considered at each step, and short-term actions may affect long-term results.

In these cases, there is a need to balance exploitation of existing knowledge (such as hiring an already known population) and exploration of sub-optimal solutions to gather more data (such as hiring populations of different backgrounds that differ from current employees).

Several studies have investigated fairness in sequential learning \citep{jabbari2017fairness,heidari2018preventing,kannan2018smoothed,joseph2016fairness,valera2018enhancing}. For example, \cite{jabbari2017fairness} study fairness in reinforcement learning and model the environment as a \textit{Markov decision process}. In their model, fairness is defined such that one action will never be preferred over another if its long-term discounted reward is lower. \cite{heidari2018preventing} define fairness as time-dependent individual fairness and require that algorithmic decisions be \textit{consistent} over time. They propose a post-process mechanism that imposes these time-dependent constraints such that two individuals that arrive during the same period and are similar in their feature dimension must be assigned similar labels.

Open challenges in this domain include the limitation of any specific time-dependent fairness definitions to the selected period of time and the effect of different discount factors.
Moreover, it is worth noting that an exploration process may be unethical on its own and impose a new type of unfairness.

In a similar line of work, researchers have investigated scenarios where feedback loops have the potential to cause amplification of bias. In these scenarios, the decisions based on the machine learning models then affect the future collected data. The risk of feedback loops is that they can introduce self-fulfilling predictions, where acting on a prediction can change the outcomes. For example, sending more police officers to an area that was predicted to be at high risk for crime will inevitably cause the arrest of more individuals in this area, and then, the prediction model will eventually further increase the risk prediction for the area \citep{ensign2018runaway}.

Note that fair sequential learning is somewhat different from another researched domain in algorithmic fairness that concerns a selection process that consists of multiple stages, such as screening candidates first by their resumes, then by their test scores and finally by interviews, where more information is gained about individuals during each stage. This field is sometimes referred to as \textit{fair pipelines} \citep{emelianov2019price,bower2017fair,hu2018short,dwork2018fairness,madras2018predict}. In these studies, fairness is revised to be considered in each stage, not only in the final stage.

\subsection{Fair Adversarial Learning}
\label{S:emerging_adversarial}
Today, adversarial learning is highly popular in the use of generative adversarial networks (GANs) \citep{goodfellow2014generative}. GANs are commonly used for the generation of simulated representative samples based on a training set. In this field of research, input data can be of various domains such as images or tabular data. In computer vision (CV), for instance, adversarial learning is used for tasks such as image generation (such as in \cite{bao2017cvae})
or modification 
(such as in \cite{antipov2017face}),
along with other CV tasks.

Nowadays, fair adversarial learning is attracting increasing attention with respect to both fair classification and the generation of fair representations. In one distressing incident, a face modifying application was exposed as racist when the app's "image filter" that was intended to change face images to more "attractive" made skin lighter \citep{Plaugic:2017:Online}.

GANs are generally constructed as a feedback loop model, starting from a generator \textit{G} that generates "fake" simulated samples and a discriminator \textit{D} (the "adversary") that determines whether the generated samples are real or fake and returns the decisions as feedback to the generator \textit{G} to improve its model.
Improving \textit{G} means enhancing its ability to generate samples that are increasingly similar to real samples in a manner that "fools" the discriminator \textit{D}, thus minimizing its ability to differentiate between real and fake samples.
Typically, both \textit{G} and \textit{D} are multi-layer neural networks.

To use GANs for fair learning, previous studies have developed different approaches. These models are often constructed as minimax optimization problems that aim at maximizing the predictor's capability to accurately predict the outcomes while minimizing the adversary's capability to predict the sensitive feature.

In one approach, it was suggested to use the feedback structure to check whether a trained classifier is fair or not and then update the model accordingly \citep{zhang2018mitigating,celis2019improved,wadsworth2018achieving}.

A different approach encourages the use of GANs to additionally learn fair representations or embedding from the training data so that it is more difficult for a subsequent classifier to distinguish which samples belong to a privileged group and which belong to an unprivileged group \citep{madras2018learning,edwards2015censoring,beutel2017data}.

Another approach \citep{xu2018fairgan,abusitta2019generative} is using GANs for generating fair \textit{synthetic} data from the initial input data and then using them to train any classifier. \cite{xu2018fairgan}, for example, use a GAN framework with one generator and two discriminators. One discriminator is trained to distinguish whether a generated sample is real or fake (as in conventional GANs), and the second identifies whether the sample belongs to the privileged or unprivileged group.

Some of these methods make efforts to also preserve semantic information of the data while learning fair representations \citep{quadrianto2019discovering,sattigeri2019fairness}.
This is essential since the interpretability and transparency of how fairness is approached in algorithmic decision-making are crucial to enhancing the understanding of decisions and trust in algorithms, and it is a paramount challenge that future research should face.

Some other papers have studied fair adversarial learning. \cite{beutel2017data} investigate the effect of input sample selection on the resulting fairness in adversarial fair learning models and show that a small balanced dataset can be effective for achieving fair representations. \cite{bose2019compositional} extend fair adversarial learning to improve fairness in \textit{graph embedding}. \cite{beutel2019putting} have raised some concerns about fair adversarial learning that should be further investigated. They argue that these models may sometimes be unstable and therefore may present some risks when using them in a production environment.

Note that other closely related problems are aimed at finding equilibrium (minimax) points in the field of \textit{game theory}, and therefore, game-theoretic schemes may also be used for solutions \citep{freund1996game,celis2019improved,agarwal2018reductions,agarwal2019fair}.

From the literature on adversarial learning, we can also note that learning fairly to predict the outcome of an unprivileged group can be thought of as learning to predict in a different domain \citep{edwards2015censoring,madras2018learning}, and therefore, notions from the field of \textit{domain adaptation} can be adopted to enhance the study of the field of \textit{algorithmic fairness}, and vice versa.

\subsection{Fair Word Embedding}

Word embedding models construct representations of words and map them to vectors (also commonly referred to as \textit{word2vec} models). Training of word embeddings is performed using raw textual data with an extremely large number of text documents and is based on the assumption that words that occur in the same contexts tend to have similar meanings. These models are primarily designed such that the embedding vectors will indicate something about the meanings and relationships between words (i.e., words with similar meanings have vectors that are close in the vector space). As such, they are broadly used in many natural language processing (NLP) applications, such as in search engines, machine translation, resume filtering, job recommendation systems, online reviews and more.

However, previous studies have shown that there are inherent biases in word embeddings (for example, \cite{bolukbasi2016man,caliskan2016semantics,zhao2018learning,brunet2019understanding}).
These studies showed that word embedding models have exhibited social biases and gender stereotypes. For instance, it has been shown that the embedding of the word "computer programmer" is closer to "male" than it is to "female". The implications of this are disturbing since these biases may affect people's lives and cause discrimination in social applications such as in job recruitment or school admissions. Another example is Microsoft's AI chat bot, named \textit{Tay}, which learned abusive language from Twitter data after only the first day of release (the bot was eventually shelved, see \cite{Hunt:2016:Online}).

A common definition of gender bias in word embedding is the measure of cosine similarity from a selected word to the words "he" and "she" (or any other two pronouns that indicate gender, such as "him"/"her" or "Mr."/"Mrs."). The difference between these two similarities may indicate the extent of bias in the embedding model. For example, consider the sentence "The \textit{CEO} raised the salary of the receptionist because \textit{he} is generous." In this sentence, "he" refers to "CEO," and further similar references in many additional sentences and texts may create a large contextual connection between these two words or other occupational nouns \citep{zhao2018learning}. Hence, even if the algorithms themselves are not biased, historical biases and norms may be embedded into the algorithm results.

To mitigate these types of biases, several studies have developed methods for debiasing the results. For example, \cite{bolukbasi2016man} suggest a post-process mechanism for removing gender bias, referred to as \textit{hard-debiasing}. Their method first identifies a \textit{gender dimension}, which is determined by a set of words that indicate gender definitions (e.g., "he"/"she"). Second, it inherently defines \textit{neutral words} (such as occupations) and then negates the projection of all of these neutral words with respect to the gender direction (so that the bias of neutral words is now zero by definition) by re-embedding the word $w$:

\begin{equation}
\vec{w}:= (\vec{w} - \vec{w_{B}})/\left \|\vec{w} - \vec{w_{B}}  \right \|
\end{equation}

\noindent where $\vec{w}$ is the embedding of the selected word and $\vec{w_{B}}$ is the projection of $w$ with respect to the gender direction.

The same paper also suggests an additional \textit{soft-debiasing} mechanism that reduces bias while still maintaining some similarity to the original embedding, thus providing a parameter that controls the trade-off between debiasing and the need to preserve information.

\cite{zhao2018learning} suggest an in-process mechanism, referred to as \textit{Gender-Neutral Global Vectors
(GN-GloVe)}, to reduce bias. Their method trains word embedding using GloVe \citep{pennington2014glove} with an altered loss function that constructs an embedding such that the protected attribute (e.g., gender) is represented in a certain dimension (a sub-vector of the embedding), which can then be ignored for debiasing. This is done by encouraging pairs of words with gender indication (e.g., "mother" and "father") to have larger distance in their gender dimension and gender-neutral words to be orthogonal to the gender direction.

\cite{brunet2019understanding} propose a pre-process mechanism for reducing bias by perturbing or removing documents during the training stage that are traced as the origin for the word embedding bias.

One challenge of these fairness mechanisms is the need for lists of words that indicate the sensitive attribute dimension and words that should be considered as neutral. \cite{bolukbasi2016man}, for example, tackle this challenge by using an initial set of gender-definitional words and train a support vector machine (SVM) to expand the list.

Let us note that the challenges of fair word embedding also affect many other downstream applications that use these word representations, for example, in \textit{coreference resolution} \citep{zhao2019gender,zhao2018gender,rudinger2018gender}, in \textit{sentence encoding} \citep{may2019measuring}, in \textit{machine translation} citep{vanmassenhove2019getting,font2019equalizing}, in \textit{language models} \citep{bordia2019identifying}, and in \textit{semantic role labeling} \citep{zhao2017men}.

Curiously, a recent study has argued that a major concern is that some of the proposed methods for removing biases in word embeddings are actually not able to remove biases but rather just "hide" them \citep{gonen2019lipstick}. They show, by a clustering illustration, that gender biases are still reflected in the embedding even after applying these methods. They additionally show that by using a support vector machine (SVM) classifier, most of the gender information can be recovered from the embedding.

Hence, it seems that existing methods as well as definitions for fair embeddings might be insufficient, and these challenges require more extensive research.

\subsection{Fair Visual Description}

The study of fairness in computer vision has recently gained extensive interest since computer vision models have been shown to produce disturbingly biased results with respect to several tasks. For example, \cite{buolamwini2018gender} have found that facial analysis models were negatively affected towards discriminating results by the under-representation of female dark-skinned faces in datasets.
\cite{kay2015unequal} show that image searches of occupations in Google's engine resulted in gender-biased results. Google's labeling application has recklessly identified black Americans as "gorillas" \citep{ Pachal:2015:Online,Simonite:2018:Online}.
Furthermore, an app that classified the attractiveness of individuals from photos turned out to be discriminative against dark skin \citep{Manthorpe:2017:Online}.

There are several previous papers in the domain of fair image classification \citep{buolamwini2018gender,quadrianto2019discovering,sattigeri2019fairness,dwork2018decoupled}. 
However, the task becomes much more complex when a fair description of images is required, such as in \textit{multi-label classification} tasks \citep{zhao2017men,van2016stereotyping, stock2017convnets,kay2015unequal}, in \textit{face attribute detection} \citep{ryu2017inclusivefacenet,karkkainen2019fairface}, or in the task of \textit{image captioning} \citep{hendricks2018women}. The last task is even more complicated than the others because of the unstructured character of the problem.

Mitigating bias in these types of problems is challenging for several reasons: First, the multimodal nature of the task requires handling fairness at both levels of natural language processing (NLP) models and computer vision (CV) models. As mentioned in a previous section, the challenges of fair word embedding also affect many other downstream applications that use these word representations (see, for example \textit{image captioning} in \cite{hendricks2018women}).
Second, the labels of the data usually depend on annotators and are not always accurate \citep{van2016stereotyping, stock2017convnets}. For example, \cite{stock2017convnets} show that human annotators fail to capture parts of the visual concepts in images and \cite{van2016stereotyping} show that the crowdsourced descriptions of the images in the \textit{Flickr30K dataset} are often based on stereotypes and prejudices of the annotators; another challenge is that the datasets are inherently unbalanced since in order to have balance - all possible co-occurrences must be balanced \citep{hendricks2018women,van2016stereotyping}

\cite{zhao2017men} consider multi-label role classification, show that some tasks display severe gender bias, and suggest a method to re-balance the resulting predictions by solving a constrained optimization problem using Lagrange relaxation. Additional constraints require the model predictions to have a similar distribution as the training set in terms of co-occurrences of gender indication and the predicted target value (this is applied to the entire corpus level since the entire corpus is needed in order to assess the frequency of occurrences).

\cite{hendricks2018women} consider the problem of fair image captioning. They propose a method to reduce gender bias by directly using a person's appearance information in the image. The method is designed to be more cautious when there is no gender information in the image. Moreover, it is constructed to function even when the distributions of genders differ between training and test sets.

\subsection{Fair Recommender Systems}

Recommender systems are prevalent in many automated and online systems and are designed to analyze users' data to provide them with personalized propositions that correspond to each user's tastes and interests. An inherent concept in recommendations is that the best items for one user may be different than those for another. Some examples of recommended items are movies, news articles, products, jobs, loans, etc. These systems have the potential to facilitate activities for both providers and consumers; however, they have also been found to exhibit fairness issues \citep{ ekstrand2017demographics,ekstrand2018exploring,burke2017multisided,burke2017balanced}.
For instance, it was shown that Google's ad-targeting algorithm had proposed higher-paying executive jobs more commonly for men than for women \citep{datta2015automated,Simonite:2015:Online}. 

Most recommender systems employ user and item similarities and are therefore prone to result in homogeneous selections that may not provide sufficient opportunities for minority populations. 

A recent paper, \cite{burke2017multisided}, notes that extending the notion of fairness from general classification tasks to recommender systems should take personalization into account. Several studies are investigating the user's perspective \citep{yao2017new, edizel2019fairecsys,ekstrand2017demographics,celis2018ranking}, where fairness is considered as recommending items equitably to different user groups (referred to as \textit{C-fairness}), such as recommending high-paying jobs to both men and women. Other studies refer to the provider's perspective \citep{lee2014fairness,ekstrand2018exploring}, where items from different providers should be recommended equally (referred to as \textit{P-fairness}), for instance, when trying to avoid market monopolistic domination.
\cite{burke2017multisided} notes that many recommender system applications involve multiple stakeholders and may therefore give rise to fairness issues for more than one group of participants simultaneously, as well as achieving fairness at a regulatory level or the level of the entire system (referred to as \textit{multisided fairness}).

It is interesting to note that \textit{P-fairness} is somewhat related to the \textit{categorical diversity} in recommender systems, requiring that recommendation lists are diverse \citep{liu2018personalizing,kunaver2017diversity}. For example, consider a hiring recommender system -- we may observe all male candidates as items of one provider and all female candidates as items of another provider. We may then ensure the equal recommendation of men and women to each of the positions.
In diversity-enhancing methods, some common measures of equality may be considered, such as the Gini coefficient that is also used in economic contexts \citep{kunaver2017diversity}. Moreover, note that an \textit{individual} definition of \textit{P-fairness}, rather than \textit{group-fairness}, may be somewhat similar to the definition of \textit{coverage} in recommender systems, requiring that each \textit{item} be recommended fairly \citep{burke2017multisided}.

\cite{surer2018multistakeholder} propose enhancing multi-stakeholder fairness using a constraint-based integer programming optimization model. The problem is computationally difficult, and hence, a following relaxation heuristic is proposed in order to solve it.
\cite{edizel2019fairecsys} suggest a post-processing mechanism that alters a fraction of the entries in the recommendation matrix so that it would be more difficult to predict the sensitive attributes from the matrix while preserving the high utility of the matrix.
The $\varepsilon$-\textit{fairness} of a recommendation matrix with respect to a certain sensitive attribute is defined as the error level when predicting the sensitive attribute using the recommendation matrix. The price of achieving $\varepsilon$-\textit{fairness} is measured by the distance between the two matrices -- the original one obtained from the prediction algorithm and the altered one obtained after the post-processing mechanism is performed.

\cite{yao2017new} and \cite{kamishima2012enhancement} suggest an in-process mechanism by including additional regularization factors in the objective functions. \cite{kamishima2012enhancement} introduce the notion of \textit{recommendation independence} to fairness-aware recommender systems. This notion requires statistical independence between recommendation results and the sensitive attribute. This means that the sensitive information will not affect the results.
\cite{yao2017new} define several notions of fairness in recommender systems. \textit{Value unfairness} measures the inconsistency in the signed estimation error across user groups. \textit{Absolute unfairness} is similar to the previous measure, but with absolute estimation. \textit{Underestimation unfairness} represents the extent to which predictions underestimate the true ratings. \textit{Overestimation unfairness} is similar to the previous measure, but with overestimation. \cite{celis2018multiwinner} and \cite{bredereck2018multiwinner} use similar notions for fairness in multiwinner voting.

One challenge in fairness-aware recommender systems is the possibility of discriminated groups based on more than one sensitive attribute \citep{edizel2019fairecsys}. It is required to first devise a definition for fairness in such cases in order to address the problem. In multi-stakeholder fairness optimization, the problems are computationally difficult \citep{surer2018multistakeholder}, so there is a need for more computational enhancements in order to work with large datasets. 
Future work may also focus on fairness in systems where there are network structures that define relationships between providers and between users (such as in \cite{bose2019compositional}).
Another possible research direction may be the incorporation of sequential notions of fairness into recommender systems through the introduction of additional time-dependent constraints. 

Note that the domain of recommender systems is also closely related to other common multi-stakeholder environments \citep{abdollahpouri2019beyond}, like \textit{resource allocation} problems such as police distribution to districts \citep{elzayn2019fair} or the allocation of aid in disaster response, where fairness is also a major concern.

\subsection{Fair Causal Learning}

Observational data collected from real world systems can mostly provide associations and correlations, rather than causal structure understandings. In contrast, causal learning relies on additional knowledge structured as a model of causes and effects.

Causal approaches may assist in enhancing fairness in several manners. For instance, by understanding causes and effects in the data, the model may assist in tackling the challenges of fairness definitions by analyzing which types of discrimination should be allowed and which should not \citep{kusner2017counterfactual,russell2017worlds}.

Another approach to improve fairness using causal reasoning is to provide an understanding of how to perform imputation of missing values or how to repair a dataset that contains sample or selection bias \citep{bareinboim2016causal,spirtes1995causal}.

Moreover, understanding a causal model may help with other ethical issues, such as defining liability and responsibility by understanding the sources of biases. This may increase the transparency and explainability of the fair models, which is also crucial for trust.

\cite{kusner2017counterfactual} have suggested a measure of causal fairness called \textit{counterfactual fairness}. It measures the extent to which it is possible to build two identical predictions $\hat{Y}$ -- one trained on the privileged group and the second trained on the unprivileged group -- using any combination of variables in the system that are not caused by the sensitive attribute. The intuition is that, in a fair model, the prediction would not change if only the sensitive attribute (and its affected variables) is changed. More precisely, a causal graph satisfies \textit{counterfactual
fairness} when the predicted label is not dependent on any descendant of the sensitive attribute.

Several studies have proposed alternative notions to \textit{counterfactual fairness}, which relax the rigid restriction on any descendant to less strict limitations. For example, the graph does not suffer from \textit{proxy discrimination} \citep{kilbertus2017avoiding} if the predicted label is not dependent on any \textit{proxy} of the sensitive attribute (a \textit{proxy} feature is a feature that can be exploited to derive the sensitive feature). Moreover, the graph does not suffer from \textit{unresolved discrimination} if the predicted label is not dependent on any \textit{resolving} variable (a \textit{resolving} variable is influenced by the sensitive feature but is accepted by practitioners as non-discriminatory).

\cite{loftus2018causal} suggest categorizing causal fairness methods according to three dimensions: individual vs. group level \citep{kilbertus2017avoiding} causal effects; explicit vs. implicit \citep{nabi2018fair} structural equations; and creating fair predictors vs. explaining and quantifying discrimination \citep{zhang2018fairness}. For example, \cite{loftus2018causal} classify \textit{counterfactual fairness} as individual, explicit and used for prediction tasks. For a broader review on causal fairness, see \cite{loftus2018causal}.

It is important to note that causal fairness models can indeed help us overcome many of the challenges encountered with respect to fair prediction tasks; however, in practice, it is difficult to obtain the correct causal model. Moreover, removing all correlated features found through a causal model may significantly compromise accuracy.

\section{Discussion and Conclusion}
\label{S:discussion}

In this paper, we presented a comprehensive and up-to-date  overview of the algorithmic fairness research field.
We started by describing the main causes of unfairness, followed by common definitions and measures of fairness, and the inevitable trade-offs between them.
We then presented fairness-enhancing mechanisms, focusing on their pros and cons, aiming at better understanding which mechanisms should be used in different scenarios.
Commonly used fairness-related datasets were then reviewed.
Lastly, we listed several emerging research sub-fields of algorithmic fairness including fair sequential learning, fair adversarial learning, fair word embedding, fair visual description, fair recommender systems and fair causal learning.

In addition to the already studied problems and the emerging ones, we identify several open challenges that should be further investigated in future research.
One major challenge stems from biases inherent in the dataset.
Such biases may arise for example, when the labeling process was performed in an already unfair manner, or if there are under-represented populations in the dataset, or in the case of systematic lack of data and in particular labels. 
Representative datasets are very difficult to achieve, and therefore, it is crucial to devise methods to overcome these issues.

Another challenge is the proliferation of definitions and measures, fairness-related datasets, and fairness-enhancing mechanisms.
It is not clear how newly proposed mechanisms should be evaluated, and in particular which measures should be considered? which datasets should be used? and which mechanisms should be used for comparison?
A closely related challenge is the difficulty in determining the balance between fairness and accuracy.
That is, what are the costs that should be assigned to each of these measures for evaluation purposes.
Future efforts should be invested in generating a benchmarking framework that will allow a more unified and standard evaluation process for fairness mechanisms.

The interpretability and transparency of how fairness is addressed by AI algorithms are another important challenge.
Such transparency is crucial to increase the understanding and trust of users in these algorithms, and in many domains, is even required by law. 
This need is further supported by several recent studies that have addressed the question of how devised mathematical notions of fairness are perceived by users \citep{srivastava2019mathematical,grgic2018human}.
It turns out that users tend to prefer the simpler notion of demographic parity, probably due to the difficulty of grasping more complex definitions of fairness.

To conclude, since the use of algorithms is expanding to all aspects of our lives, demanding that automated decisions be more ethical and fair is inevitable.
We should aspire to not only develop fairer algorithms, but also to design procedures to reduce biases in the data.
Such procedures may rely for example on integrating both humans and algorithms in the decision pipeline.
However,  
thus far, it seems that biased algorithms are easier to fix than biased humans or procedures \citep{Mullainathan:2019:Online}.

\begin{acks}
This paper was partially supported by the Koret foundation grant for Smart Cities and Digital
Living 2030.
\end{acks}

\bibliographystyle{ACM-Reference-Format}
\bibliography{mybibacm}


\begin{thebibliography}{152}


\ifx \showCODEN    \undefined \def \showCODEN     #1{\unskip}     \fi
\ifx \showDOI      \undefined \def \showDOI       #1{#1}\fi
\ifx \showISBNx    \undefined \def \showISBNx     #1{\unskip}     \fi
\ifx \showISBNxiii \undefined \def \showISBNxiii  #1{\unskip}     \fi
\ifx \showISSN     \undefined \def \showISSN      #1{\unskip}     \fi
\ifx \showLCCN     \undefined \def \showLCCN      #1{\unskip}     \fi
\ifx \shownote     \undefined \def \shownote      #1{#1}          \fi
\ifx \showarticletitle \undefined \def \showarticletitle #1{#1}   \fi
\ifx \showURL      \undefined \def \showURL       {\relax}        \fi
\providecommand\bibfield[2]{#2}
\providecommand\bibinfo[2]{#2}
\providecommand\natexlab[1]{#1}
\providecommand\showeprint[2][]{arXiv:#2}

\bibitem[\protect\citeauthoryear{??}{com}{2012}]%
        {compas}
 \bibinfo{year}{2012}\natexlab{}.
\newblock \bibinfo{title}{Practitioners Guide to COMPAS}.
\newblock
\newblock
\urldef\tempurl%
\url{http://www.northpointeinc.com/files/technical\_documents/FieldGuide2\_081412.pdf}
\showURL{%
\tempurl}


\bibitem[\protect\citeauthoryear{Abdollahpouri, Adomavicius, Burke, Guy,
  Jannach, Kamishima, Krasnodebski, and Pizzato}{Abdollahpouri
  et~al\mbox{.}}{2019}]%
        {abdollahpouri2019beyond}
\bibfield{author}{\bibinfo{person}{Himan Abdollahpouri},
  \bibinfo{person}{Gediminas Adomavicius}, \bibinfo{person}{Robin Burke},
  \bibinfo{person}{Ido Guy}, \bibinfo{person}{Dietmar Jannach},
  \bibinfo{person}{Toshihiro Kamishima}, \bibinfo{person}{Jan Krasnodebski},
  {and} \bibinfo{person}{Luiz Pizzato}.} \bibinfo{year}{2019}\natexlab{}.
\newblock \bibinfo{title}{Beyond Personalization: Research Directions in
  Multistakeholder Recommendation}.
\newblock
\newblock


\bibitem[\protect\citeauthoryear{Abusitta, A{\"\i}meur, and Wahab}{Abusitta
  et~al\mbox{.}}{2019}]%
        {abusitta2019generative}
\bibfield{author}{\bibinfo{person}{Adel Abusitta}, \bibinfo{person}{Esma
  A{\"\i}meur}, {and} \bibinfo{person}{Omar~Abdel Wahab}.}
  \bibinfo{year}{2019}\natexlab{}.
\newblock \bibinfo{title}{Generative Adversarial Networks for Mitigating Biases
  in Machine Learning Systems}.
\newblock
\newblock


\bibitem[\protect\citeauthoryear{Agarwal, Beygelzimer, Dudik, Langford, and
  Wallach}{Agarwal et~al\mbox{.}}{2018}]%
        {agarwal2018reductions}
\bibfield{author}{\bibinfo{person}{Alekh Agarwal}, \bibinfo{person}{Alina
  Beygelzimer}, \bibinfo{person}{Miroslav Dudik}, \bibinfo{person}{John
  Langford}, {and} \bibinfo{person}{Hanna Wallach}.}
  \bibinfo{year}{2018}\natexlab{}.
\newblock \showarticletitle{A Reductions Approach to Fair Classification}. In
  \bibinfo{booktitle}{\emph{International Conference on Machine Learning}}.
  \bibinfo{pages}{60--69}.
\newblock


\bibitem[\protect\citeauthoryear{Agarwal, Dudik, and Wu}{Agarwal
  et~al\mbox{.}}{2019}]%
        {agarwal2019fair}
\bibfield{author}{\bibinfo{person}{Alekh Agarwal}, \bibinfo{person}{Miroslav
  Dudik}, {and} \bibinfo{person}{Zhiwei~Steven Wu}.}
  \bibinfo{year}{2019}\natexlab{}.
\newblock \showarticletitle{Fair Regression: Quantitative Definitions and
  Reduction-Based Algorithms}. In \bibinfo{booktitle}{\emph{International
  Conference on Machine Learning}}. \bibinfo{pages}{120--129}.
\newblock


\bibitem[\protect\citeauthoryear{Angwin}{Angwin}{2016}]%
        {Angwin:2016:Online}
\bibfield{author}{\bibinfo{person}{Julia Angwin}.}
  \bibinfo{year}{2016}\natexlab{}.
\newblock \bibinfo{title}{Machine Bias — ProPublica}.
\newblock
\newblock
\urldef\tempurl%
\url{https://www.propublica.org/article/machine-bias-risk-assessments-in-criminal-sentencing}
\showURL{%
\tempurl}


\bibitem[\protect\citeauthoryear{Antipov, Baccouche, and Dugelay}{Antipov
  et~al\mbox{.}}{2017}]%
        {antipov2017face}
\bibfield{author}{\bibinfo{person}{Grigory Antipov}, \bibinfo{person}{Moez
  Baccouche}, {and} \bibinfo{person}{Jean-Luc Dugelay}.}
  \bibinfo{year}{2017}\natexlab{}.
\newblock \showarticletitle{Face aging with conditional generative adversarial
  networks}. In \bibinfo{booktitle}{\emph{2017 IEEE International Conference on
  Image Processing (ICIP)}}. \bibinfo{publisher}{IEEE},
  \bibinfo{pages}{2089--2093}.
\newblock


\bibitem[\protect\citeauthoryear{Backurs, Indyk, Onak, Schieber, Vakilian, and
  Wagner}{Backurs et~al\mbox{.}}{2019}]%
        {backurs2019scalable}
\bibfield{author}{\bibinfo{person}{Arturs Backurs}, \bibinfo{person}{Piotr
  Indyk}, \bibinfo{person}{Krzysztof Onak}, \bibinfo{person}{Baruch Schieber},
  \bibinfo{person}{Ali Vakilian}, {and} \bibinfo{person}{Tal Wagner}.}
  \bibinfo{year}{2019}\natexlab{}.
\newblock \showarticletitle{Scalable Fair Clustering}. In
  \bibinfo{booktitle}{\emph{International Conference on Machine Learning}}.
  \bibinfo{pages}{405--413}.
\newblock


\bibitem[\protect\citeauthoryear{Bao, Chen, Wen, Li, and Hua}{Bao
  et~al\mbox{.}}{2017}]%
        {bao2017cvae}
\bibfield{author}{\bibinfo{person}{Jianmin Bao}, \bibinfo{person}{Dong Chen},
  \bibinfo{person}{Fang Wen}, \bibinfo{person}{Houqiang Li}, {and}
  \bibinfo{person}{Gang Hua}.} \bibinfo{year}{2017}\natexlab{}.
\newblock \showarticletitle{CVAE-GAN: fine-grained image generation through
  asymmetric training}. In \bibinfo{booktitle}{\emph{Proceedings of the IEEE
  International Conference on Computer Vision}}. \bibinfo{pages}{2745--2754}.
\newblock


\bibitem[\protect\citeauthoryear{Bareinboim and Pearl}{Bareinboim and
  Pearl}{2016}]%
        {bareinboim2016causal}
\bibfield{author}{\bibinfo{person}{Elias Bareinboim} {and}
  \bibinfo{person}{Judea Pearl}.} \bibinfo{year}{2016}\natexlab{}.
\newblock \showarticletitle{Causal inference and the data-fusion problem}.
\newblock \bibinfo{journal}{\emph{Proceedings of the National Academy of
  Sciences}} \bibinfo{volume}{113}, \bibinfo{number}{27}
  (\bibinfo{year}{2016}), \bibinfo{pages}{7345--7352}.
\newblock


\bibitem[\protect\citeauthoryear{Barocas and Selbst}{Barocas and
  Selbst}{2016}]%
        {barocas2016big}
\bibfield{author}{\bibinfo{person}{Solon Barocas} {and}
  \bibinfo{person}{Andrew~D Selbst}.} \bibinfo{year}{2016}\natexlab{}.
\newblock \showarticletitle{Big data's disparate impact}.
\newblock \bibinfo{journal}{\emph{Calif. L. Rev.}}  \bibinfo{volume}{104}
  (\bibinfo{year}{2016}), \bibinfo{pages}{671}.
\newblock


\bibitem[\protect\citeauthoryear{Bechavod and Ligett}{Bechavod and
  Ligett}{2017a}]%
        {bechavod2017learning}
\bibfield{author}{\bibinfo{person}{Yahav Bechavod} {and}
  \bibinfo{person}{Katrina Ligett}.} \bibinfo{year}{2017}\natexlab{a}.
\newblock \bibinfo{title}{Learning fair classifiers: A regularization-inspired
  approach}.
\newblock , \bibinfo{numpages}{1733--1782}~pages.
\newblock


\bibitem[\protect\citeauthoryear{Bechavod and Ligett}{Bechavod and
  Ligett}{2017b}]%
        {bechavod2017penalizing}
\bibfield{author}{\bibinfo{person}{Yahav Bechavod} {and}
  \bibinfo{person}{Katrina Ligett}.} \bibinfo{year}{2017}\natexlab{b}.
\newblock \bibinfo{title}{Penalizing unfairness in binary classification}.
\newblock
\newblock


\bibitem[\protect\citeauthoryear{Berk, Heidari, Jabbari, Joseph, Kearns,
  Morgenstern, Neel, and Roth}{Berk et~al\mbox{.}}{2017}]%
        {berk2017convex}
\bibfield{author}{\bibinfo{person}{Richard Berk}, \bibinfo{person}{Hoda
  Heidari}, \bibinfo{person}{Shahin Jabbari}, \bibinfo{person}{Matthew Joseph},
  \bibinfo{person}{Michael Kearns}, \bibinfo{person}{Jamie Morgenstern},
  \bibinfo{person}{Seth Neel}, {and} \bibinfo{person}{Aaron Roth}.}
  \bibinfo{year}{2017}\natexlab{}.
\newblock \bibinfo{title}{A convex framework for fair regression}.
\newblock
\newblock


\bibitem[\protect\citeauthoryear{Berk, Heidari, Jabbari, Kearns, and Roth}{Berk
  et~al\mbox{.}}{2018}]%
        {berk2018fairness}
\bibfield{author}{\bibinfo{person}{Richard Berk}, \bibinfo{person}{Hoda
  Heidari}, \bibinfo{person}{Shahin Jabbari}, \bibinfo{person}{Michael Kearns},
  {and} \bibinfo{person}{Aaron Roth}.} \bibinfo{year}{2018}\natexlab{}.
\newblock \showarticletitle{Fairness in criminal justice risk assessments: The
  state of the art}.
\newblock \bibinfo{journal}{\emph{Sociological Methods \& Research}}
  (\bibinfo{year}{2018}), \bibinfo{pages}{0049124118782533}.
\newblock


\bibitem[\protect\citeauthoryear{Beutel, Chen, Doshi, Qian, Woodruff, Luu,
  Kreitmann, Bischof, and Chi}{Beutel et~al\mbox{.}}{2019}]%
        {beutel2019putting}
\bibfield{author}{\bibinfo{person}{Alex Beutel}, \bibinfo{person}{Jilin Chen},
  \bibinfo{person}{Tulsee Doshi}, \bibinfo{person}{Hai Qian},
  \bibinfo{person}{Allison Woodruff}, \bibinfo{person}{Christine Luu},
  \bibinfo{person}{Pierre Kreitmann}, \bibinfo{person}{Jonathan Bischof}, {and}
  \bibinfo{person}{Ed~H Chi}.} \bibinfo{year}{2019}\natexlab{}.
\newblock \bibinfo{title}{Putting fairness principles into practice:
  Challenges, metrics, and improvements}.
\newblock
\newblock


\bibitem[\protect\citeauthoryear{Beutel, Chen, Zhao, and Chi}{Beutel
  et~al\mbox{.}}{2017}]%
        {beutel2017data}
\bibfield{author}{\bibinfo{person}{Alex Beutel}, \bibinfo{person}{Jilin Chen},
  \bibinfo{person}{Zhe Zhao}, {and} \bibinfo{person}{Ed~H Chi}.}
  \bibinfo{year}{2017}\natexlab{}.
\newblock \bibinfo{title}{Data decisions and theoretical implications when
  adversarially learning fair representations}.
\newblock
\newblock


\bibitem[\protect\citeauthoryear{Bolukbasi, Chang, Zou, Saligrama, and
  Kalai}{Bolukbasi et~al\mbox{.}}{2016}]%
        {bolukbasi2016man}
\bibfield{author}{\bibinfo{person}{Tolga Bolukbasi}, \bibinfo{person}{Kai-Wei
  Chang}, \bibinfo{person}{James~Y Zou}, \bibinfo{person}{Venkatesh Saligrama},
  {and} \bibinfo{person}{Adam~T Kalai}.} \bibinfo{year}{2016}\natexlab{}.
\newblock \showarticletitle{Man is to computer programmer as woman is to
  homemaker? debiasing word embeddings}. In \bibinfo{booktitle}{\emph{Advances
  in neural information processing systems}}. \bibinfo{pages}{4349--4357}.
\newblock


\bibitem[\protect\citeauthoryear{Bordia and Bowman}{Bordia and Bowman}{2019}]%
        {bordia2019identifying}
\bibfield{author}{\bibinfo{person}{Shikha Bordia} {and} \bibinfo{person}{Samuel
  Bowman}.} \bibinfo{year}{2019}\natexlab{}.
\newblock \showarticletitle{Identifying and Reducing Gender Bias in Word-Level
  Language Models}. In \bibinfo{booktitle}{\emph{Proceedings of the 2019
  Conference of the North American Chapter of the Association for Computational
  Linguistics: Student Research Workshop}}. \bibinfo{pages}{7--15}.
\newblock


\bibitem[\protect\citeauthoryear{Bose and Hamilton}{Bose and Hamilton}{2019}]%
        {bose2019compositional}
\bibfield{author}{\bibinfo{person}{Avishek Bose} {and} \bibinfo{person}{William
  Hamilton}.} \bibinfo{year}{2019}\natexlab{}.
\newblock \showarticletitle{Compositional Fairness Constraints for Graph
  Embeddings}. In \bibinfo{booktitle}{\emph{International Conference on Machine
  Learning}}. \bibinfo{pages}{715--724}.
\newblock


\bibitem[\protect\citeauthoryear{Bower, Kitchen, Niss, Strauss, Vargas, and
  Venkatasubramanian}{Bower et~al\mbox{.}}{2017}]%
        {bower2017fair}
\bibfield{author}{\bibinfo{person}{Amanda Bower}, \bibinfo{person}{Sarah~N
  Kitchen}, \bibinfo{person}{Laura Niss}, \bibinfo{person}{Martin~J Strauss},
  \bibinfo{person}{Alexander Vargas}, {and} \bibinfo{person}{Suresh
  Venkatasubramanian}.} \bibinfo{year}{2017}\natexlab{}.
\newblock \bibinfo{title}{Fair pipelines}.
\newblock
\newblock


\bibitem[\protect\citeauthoryear{Bredereck, Faliszewski, Igarashi, Lackner, and
  Skowron}{Bredereck et~al\mbox{.}}{2018}]%
        {bredereck2018multiwinner}
\bibfield{author}{\bibinfo{person}{Robert Bredereck}, \bibinfo{person}{Piotr
  Faliszewski}, \bibinfo{person}{Ayumi Igarashi}, \bibinfo{person}{Martin
  Lackner}, {and} \bibinfo{person}{Piotr Skowron}.}
  \bibinfo{year}{2018}\natexlab{}.
\newblock \showarticletitle{Multiwinner elections with diversity constraints}.
  In \bibinfo{booktitle}{\emph{Thirty-Second AAAI Conference on Artificial
  Intelligence}}.
\newblock


\bibitem[\protect\citeauthoryear{Brierley, Vogel, and Axelrod}{Brierley
  et~al\mbox{.}}{2011}]%
        {brierley2011heritage}
\bibfield{author}{\bibinfo{person}{Phil Brierley}, \bibinfo{person}{David
  Vogel}, {and} \bibinfo{person}{Randy Axelrod}.}
  \bibinfo{year}{2011}\natexlab{}.
\newblock \bibinfo{title}{Heritage Provider Network Health Prize Round 1
  Milestone Prize: How we did it--Team ‘Market Makers’}.
\newblock
\newblock


\bibitem[\protect\citeauthoryear{Brunet, Alkalay-Houlihan, Anderson, and
  Zemel}{Brunet et~al\mbox{.}}{2019}]%
        {brunet2019understanding}
\bibfield{author}{\bibinfo{person}{Marc-Etienne Brunet},
  \bibinfo{person}{Colleen Alkalay-Houlihan}, \bibinfo{person}{Ashton
  Anderson}, {and} \bibinfo{person}{Richard Zemel}.}
  \bibinfo{year}{2019}\natexlab{}.
\newblock \showarticletitle{Understanding the Origins of Bias in Word
  Embeddings}. In \bibinfo{booktitle}{\emph{International Conference on Machine
  Learning}}. \bibinfo{pages}{803--811}.
\newblock


\bibitem[\protect\citeauthoryear{Buolamwini and Gebru}{Buolamwini and
  Gebru}{2018}]%
        {buolamwini2018gender}
\bibfield{author}{\bibinfo{person}{Joy Buolamwini} {and}
  \bibinfo{person}{Timnit Gebru}.} \bibinfo{year}{2018}\natexlab{}.
\newblock \showarticletitle{Gender shades: Intersectional accuracy disparities
  in commercial gender classification}. In \bibinfo{booktitle}{\emph{Conference
  on fairness, accountability and transparency}}. \bibinfo{pages}{77--91}.
\newblock


\bibitem[\protect\citeauthoryear{Burke}{Burke}{2017}]%
        {burke2017multisided}
\bibfield{author}{\bibinfo{person}{Robin Burke}.}
  \bibinfo{year}{2017}\natexlab{}.
\newblock \bibinfo{title}{Multisided Fairness for Recommendation}.
\newblock
\newblock


\bibitem[\protect\citeauthoryear{Burke, Sonboli, Mansoury, and
  Ordo{\~n}ez-Gauger}{Burke et~al\mbox{.}}{2017}]%
        {burke2017balanced}
\bibfield{author}{\bibinfo{person}{Robin Burke}, \bibinfo{person}{Nasim
  Sonboli}, \bibinfo{person}{Masoud Mansoury}, {and} \bibinfo{person}{Aldo
  Ordo{\~n}ez-Gauger}.} \bibinfo{year}{2017}\natexlab{}.
\newblock \showarticletitle{Balanced Neighborhoods for Fairness-aware
  Collaborative Recommendation}.
\newblock \bibinfo{journal}{\emph{Proceedings of ACM FATRec Workshop, Como,
  Italy}} (\bibinfo{year}{2017}).
\newblock


\bibitem[\protect\citeauthoryear{Calders and Verwer}{Calders and
  Verwer}{2010}]%
        {calders2010three}
\bibfield{author}{\bibinfo{person}{Toon Calders} {and} \bibinfo{person}{Sicco
  Verwer}.} \bibinfo{year}{2010}\natexlab{}.
\newblock \showarticletitle{Three naive Bayes approaches for
  discrimination-free classification}.
\newblock \bibinfo{journal}{\emph{Data Mining and Knowledge Discovery}}
  \bibinfo{volume}{21}, \bibinfo{number}{2} (\bibinfo{year}{2010}),
  \bibinfo{pages}{277--292}.
\newblock


\bibitem[\protect\citeauthoryear{Caliskan, Bryson, and Narayanan}{Caliskan
  et~al\mbox{.}}{2016}]%
        {caliskan2016semantics}
\bibfield{author}{\bibinfo{person}{Aylin Caliskan}, \bibinfo{person}{Joanna~J
  Bryson}, {and} \bibinfo{person}{Arvind Narayanan}.}
  \bibinfo{year}{2016}\natexlab{}.
\newblock \showarticletitle{Semantics derived automatically from language
  corpora necessarily contain human biases}.
\newblock \bibinfo{journal}{\emph{CoRR, abs/1608.07187}}
  (\bibinfo{year}{2016}).
\newblock


\bibitem[\protect\citeauthoryear{Calmon, Wei, Vinzamuri, Ramamurthy, and
  Varshney}{Calmon et~al\mbox{.}}{2017}]%
        {calmon2017optimized}
\bibfield{author}{\bibinfo{person}{Flavio Calmon}, \bibinfo{person}{Dennis
  Wei}, \bibinfo{person}{Bhanukiran Vinzamuri},
  \bibinfo{person}{Karthikeyan~Natesan Ramamurthy}, {and}
  \bibinfo{person}{Kush~R Varshney}.} \bibinfo{year}{2017}\natexlab{}.
\newblock \showarticletitle{Optimized pre-processing for discrimination
  prevention}. In \bibinfo{booktitle}{\emph{Advances in Neural Information
  Processing Systems}}. \bibinfo{pages}{3992--4001}.
\newblock


\bibitem[\protect\citeauthoryear{Celis, Huang, and Vishnoi}{Celis
  et~al\mbox{.}}{2018a}]%
        {celis2018multiwinner}
\bibfield{author}{\bibinfo{person}{L~Elisa Celis}, \bibinfo{person}{Lingxiao
  Huang}, {and} \bibinfo{person}{Nisheeth~K Vishnoi}.}
  \bibinfo{year}{2018}\natexlab{a}.
\newblock \showarticletitle{Multiwinner voting with fairness constraints}. In
  \bibinfo{booktitle}{\emph{Proceedings of the 27th International Joint
  Conference on Artificial Intelligence}}. \bibinfo{publisher}{AAAI Press},
  \bibinfo{pages}{144--151}.
\newblock


\bibitem[\protect\citeauthoryear{Celis and Keswani}{Celis and Keswani}{2019}]%
        {celis2019improved}
\bibfield{author}{\bibinfo{person}{L~Elisa Celis} {and} \bibinfo{person}{Vijay
  Keswani}.} \bibinfo{year}{2019}\natexlab{}.
\newblock \bibinfo{title}{Improved Adversarial Learning for Fair
  Classification}.
\newblock
\newblock


\bibitem[\protect\citeauthoryear{Celis, Straszak, and Vishnoi}{Celis
  et~al\mbox{.}}{2018b}]%
        {celis2018ranking}
\bibfield{author}{\bibinfo{person}{L~Elisa Celis}, \bibinfo{person}{Damian
  Straszak}, {and} \bibinfo{person}{Nisheeth~K Vishnoi}.}
  \bibinfo{year}{2018}\natexlab{b}.
\newblock \showarticletitle{Ranking with Fairness Constraints}. In
  \bibinfo{booktitle}{\emph{45th International Colloquium on Automata,
  Languages, and Programming (ICALP 2018)}}. \bibinfo{publisher}{Schloss
  Dagstuhl-Leibniz-Zentrum fuer Informatik}.
\newblock


\bibitem[\protect\citeauthoryear{Center}{Center}{2015}]%
        {minnesota2015integrated}
\bibfield{author}{\bibinfo{person}{Minnesota~Population Center}.}
  \bibinfo{year}{2015}\natexlab{}.
\newblock \bibinfo{title}{Integrated Public Use Microdata Series,
  International: Version 6.4 [The Dutch Virtual Census of 2001]}.
\newblock
\newblock
\urldef\tempurl%
\url{https://doi.org/10.18128/D020.V6.4}
\showDOI{\tempurl}


\bibitem[\protect\citeauthoryear{Chierichetti, Kumar, Lattanzi, and
  Vassilvitskii}{Chierichetti et~al\mbox{.}}{2017}]%
        {chierichetti2017fair}
\bibfield{author}{\bibinfo{person}{Flavio Chierichetti}, \bibinfo{person}{Ravi
  Kumar}, \bibinfo{person}{Silvio Lattanzi}, {and} \bibinfo{person}{Sergei
  Vassilvitskii}.} \bibinfo{year}{2017}\natexlab{}.
\newblock \showarticletitle{Fair clustering through fairlets}. In
  \bibinfo{booktitle}{\emph{Advances in Neural Information Processing
  Systems}}. \bibinfo{pages}{5029--5037}.
\newblock


\bibitem[\protect\citeauthoryear{Chouldechova}{Chouldechova}{2017}]%
        {chouldechova2017fair}
\bibfield{author}{\bibinfo{person}{Alexandra Chouldechova}.}
  \bibinfo{year}{2017}\natexlab{}.
\newblock \showarticletitle{Fair prediction with disparate impact: A study of
  bias in recidivism prediction instruments}.
\newblock \bibinfo{journal}{\emph{Big data}} \bibinfo{volume}{5},
  \bibinfo{number}{2} (\bibinfo{year}{2017}), \bibinfo{pages}{153--163}.
\newblock


\bibitem[\protect\citeauthoryear{Chouldechova and Roth}{Chouldechova and
  Roth}{2018}]%
        {chouldechova2018frontiers}
\bibfield{author}{\bibinfo{person}{Alexandra Chouldechova} {and}
  \bibinfo{person}{Aaron Roth}.} \bibinfo{year}{2018}\natexlab{}.
\newblock \bibinfo{title}{The frontiers of fairness in machine learning}.
\newblock
\newblock


\bibitem[\protect\citeauthoryear{Corbett-Davies and Goel}{Corbett-Davies and
  Goel}{2018}]%
        {corbett2018measure}
\bibfield{author}{\bibinfo{person}{Sam Corbett-Davies} {and}
  \bibinfo{person}{Sharad Goel}.} \bibinfo{year}{2018}\natexlab{}.
\newblock \bibinfo{title}{The measure and mismeasure of fairness: A critical
  review of fair machine learning}.
\newblock
\newblock


\bibitem[\protect\citeauthoryear{Corbett-Davies, Pierson, Feller, Goel, and
  Huq}{Corbett-Davies et~al\mbox{.}}{2017}]%
        {corbett2017algorithmic}
\bibfield{author}{\bibinfo{person}{Sam Corbett-Davies}, \bibinfo{person}{Emma
  Pierson}, \bibinfo{person}{Avi Feller}, \bibinfo{person}{Sharad Goel}, {and}
  \bibinfo{person}{Aziz Huq}.} \bibinfo{year}{2017}\natexlab{}.
\newblock \showarticletitle{Algorithmic decision making and the cost of
  fairness}. In \bibinfo{booktitle}{\emph{Proceedings of the 23rd ACM SIGKDD
  International Conference on Knowledge Discovery and Data Mining}}.
  \bibinfo{publisher}{ACM}, \bibinfo{pages}{797--806}.
\newblock


\bibitem[\protect\citeauthoryear{Dastin}{Dastin}{2018}]%
        {Dastin:2018:Online}
\bibfield{author}{\bibinfo{person}{Jeffrey Dastin}.}
  \bibinfo{year}{2018}\natexlab{}.
\newblock \bibinfo{title}{Amazon scraps secret AI recruiting tool that showed
  bias against women}.
\newblock
\newblock
\urldef\tempurl%
\url{https://www.reuters.com/article/us-amazon-com-jobs-automation-insight/amazon-scraps-secret-ai-recruiting-tool-that-showed-bias-against-women-idUSKCN1MK08G}
\showURL{%
\tempurl}


\bibitem[\protect\citeauthoryear{Datta, Tschantz, and Datta}{Datta
  et~al\mbox{.}}{2015}]%
        {datta2015automated}
\bibfield{author}{\bibinfo{person}{Amit Datta}, \bibinfo{person}{Michael~Carl
  Tschantz}, {and} \bibinfo{person}{Anupam Datta}.}
  \bibinfo{year}{2015}\natexlab{}.
\newblock \showarticletitle{Automated experiments on ad privacy settings}.
\newblock \bibinfo{journal}{\emph{Proceedings on privacy enhancing
  technologies}} \bibinfo{volume}{2015}, \bibinfo{number}{1}
  (\bibinfo{year}{2015}), \bibinfo{pages}{92--112}.
\newblock


\bibitem[\protect\citeauthoryear{Deng, Dong, Socher, Li, Li, and Fei-Fei}{Deng
  et~al\mbox{.}}{2009}]%
        {deng2009imagenet}
\bibfield{author}{\bibinfo{person}{Jia Deng}, \bibinfo{person}{Wei Dong},
  \bibinfo{person}{Richard Socher}, \bibinfo{person}{Li-Jia Li},
  \bibinfo{person}{Kai Li}, {and} \bibinfo{person}{Li Fei-Fei}.}
  \bibinfo{year}{2009}\natexlab{}.
\newblock \showarticletitle{Imagenet: A large-scale hierarchical image
  database}. In \bibinfo{booktitle}{\emph{2009 IEEE conference on computer
  vision and pattern recognition}}. \bibinfo{publisher}{Ieee},
  \bibinfo{pages}{248--255}.
\newblock


\bibitem[\protect\citeauthoryear{Dua and Graff}{Dua and Graff}{2017}]%
        {Dua:2019}
\bibfield{author}{\bibinfo{person}{Dheeru Dua} {and} \bibinfo{person}{Casey
  Graff}.} \bibinfo{year}{2017}\natexlab{}.
\newblock \bibinfo{title}{{UCI} Machine Learning Repository}.
\newblock
\newblock
\urldef\tempurl%
\url{http://archive.ics.uci.edu/ml}
\showURL{%
\tempurl}


\bibitem[\protect\citeauthoryear{Dwork, Hardt, Pitassi, Reingold, and
  Zemel}{Dwork et~al\mbox{.}}{2012}]%
        {dwork2012fairness}
\bibfield{author}{\bibinfo{person}{Cynthia Dwork}, \bibinfo{person}{Moritz
  Hardt}, \bibinfo{person}{Toniann Pitassi}, \bibinfo{person}{Omer Reingold},
  {and} \bibinfo{person}{Richard Zemel}.} \bibinfo{year}{2012}\natexlab{}.
\newblock \showarticletitle{Fairness through awareness}. In
  \bibinfo{booktitle}{\emph{Proceedings of the 3rd innovations in theoretical
  computer science conference}}. \bibinfo{publisher}{ACM},
  \bibinfo{pages}{214--226}.
\newblock


\bibitem[\protect\citeauthoryear{Dwork and Ilvento}{Dwork and Ilvento}{2018}]%
        {dwork2018fairness}
\bibfield{author}{\bibinfo{person}{Cynthia Dwork} {and}
  \bibinfo{person}{Christina Ilvento}.} \bibinfo{year}{2018}\natexlab{}.
\newblock \showarticletitle{Fairness Under Composition}. In
  \bibinfo{booktitle}{\emph{10th Innovations in Theoretical Computer Science
  Conference (ITCS 2019)}}. \bibinfo{publisher}{Schloss
  Dagstuhl-Leibniz-Zentrum fuer Informatik}.
\newblock


\bibitem[\protect\citeauthoryear{Dwork, Immorlica, Kalai, and Leiserson}{Dwork
  et~al\mbox{.}}{2018}]%
        {dwork2018decoupled}
\bibfield{author}{\bibinfo{person}{Cynthia Dwork}, \bibinfo{person}{Nicole
  Immorlica}, \bibinfo{person}{Adam~Tauman Kalai}, {and} \bibinfo{person}{Max
  Leiserson}.} \bibinfo{year}{2018}\natexlab{}.
\newblock \showarticletitle{Decoupled classifiers for group-fair and efficient
  machine learning}. In \bibinfo{booktitle}{\emph{Conference on Fairness,
  Accountability and Transparency}}. \bibinfo{pages}{119--133}.
\newblock


\bibitem[\protect\citeauthoryear{Edizel, Bonchi, Hajian, Panisson, and
  Tassa}{Edizel et~al\mbox{.}}{2019}]%
        {edizel2019fairecsys}
\bibfield{author}{\bibinfo{person}{Bora Edizel}, \bibinfo{person}{Francesco
  Bonchi}, \bibinfo{person}{Sara Hajian}, \bibinfo{person}{Andr{\'e} Panisson},
  {and} \bibinfo{person}{Tamir Tassa}.} \bibinfo{year}{2019}\natexlab{}.
\newblock \showarticletitle{FaiRecSys: mitigating algorithmic bias in
  recommender systems}.
\newblock \bibinfo{journal}{\emph{International Journal of Data Science and
  Analytics}} (\bibinfo{year}{2019}), \bibinfo{pages}{1--17}.
\newblock


\bibitem[\protect\citeauthoryear{Edwards and Storkey}{Edwards and
  Storkey}{2015}]%
        {edwards2015censoring}
\bibfield{author}{\bibinfo{person}{Harrison Edwards} {and}
  \bibinfo{person}{Amos Storkey}.} \bibinfo{year}{2015}\natexlab{}.
\newblock \bibinfo{title}{Censoring representations with an adversary}.
\newblock
\newblock


\bibitem[\protect\citeauthoryear{Ekstrand and Pera}{Ekstrand and Pera}{2017}]%
        {ekstrand2017demographics}
\bibfield{author}{\bibinfo{person}{Michael~D Ekstrand} {and}
  \bibinfo{person}{Maria~Soledad Pera}.} \bibinfo{year}{2017}\natexlab{}.
\newblock \showarticletitle{The demographics of cool}.
\newblock \bibinfo{journal}{\emph{Poster Proceedings at ACM RecSys. ACM, Como,
  Italy}} (\bibinfo{year}{2017}).
\newblock


\bibitem[\protect\citeauthoryear{Ekstrand, Tian, Kazi, Mehrpouyan, and
  Kluver}{Ekstrand et~al\mbox{.}}{2018}]%
        {ekstrand2018exploring}
\bibfield{author}{\bibinfo{person}{Michael~D Ekstrand}, \bibinfo{person}{Mucun
  Tian}, \bibinfo{person}{Mohammed R~Imran Kazi}, \bibinfo{person}{Hoda
  Mehrpouyan}, {and} \bibinfo{person}{Daniel Kluver}.}
  \bibinfo{year}{2018}\natexlab{}.
\newblock \showarticletitle{Exploring author gender in book rating and
  recommendation}. In \bibinfo{booktitle}{\emph{Proceedings of the 12th ACM
  Conference on Recommender Systems}}. \bibinfo{publisher}{ACM},
  \bibinfo{pages}{242--250}.
\newblock


\bibitem[\protect\citeauthoryear{Elzayn, Jabbari, Jung, Kearns, Neel, Roth, and
  Schutzman}{Elzayn et~al\mbox{.}}{2019}]%
        {elzayn2019fair}
\bibfield{author}{\bibinfo{person}{Hadi Elzayn}, \bibinfo{person}{Shahin
  Jabbari}, \bibinfo{person}{Christopher Jung}, \bibinfo{person}{Michael
  Kearns}, \bibinfo{person}{Seth Neel}, \bibinfo{person}{Aaron Roth}, {and}
  \bibinfo{person}{Zachary Schutzman}.} \bibinfo{year}{2019}\natexlab{}.
\newblock \showarticletitle{Fair algorithms for learning in allocation
  problems}. In \bibinfo{booktitle}{\emph{Proceedings of the Conference on
  Fairness, Accountability, and Transparency}}. \bibinfo{publisher}{ACM},
  \bibinfo{pages}{170--179}.
\newblock


\bibitem[\protect\citeauthoryear{Emelianov, Arvanitakis, Gast, Gummadi, and
  Loiseau}{Emelianov et~al\mbox{.}}{2019}]%
        {emelianov2019price}
\bibfield{author}{\bibinfo{person}{Vitalii Emelianov}, \bibinfo{person}{George
  Arvanitakis}, \bibinfo{person}{Nicolas Gast}, \bibinfo{person}{Krishna
  Gummadi}, {and} \bibinfo{person}{Patrick Loiseau}.}
  \bibinfo{year}{2019}\natexlab{}.
\newblock \bibinfo{title}{The Price of Local Fairness in Multistage Selection}.
\newblock
\newblock


\bibitem[\protect\citeauthoryear{Ensign, Friedler, Neville, Scheidegger, and
  Venkatasubramanian}{Ensign et~al\mbox{.}}{2018}]%
        {ensign2018runaway}
\bibfield{author}{\bibinfo{person}{Danielle Ensign}, \bibinfo{person}{Sorelle~A
  Friedler}, \bibinfo{person}{Scott Neville}, \bibinfo{person}{Carlos
  Scheidegger}, {and} \bibinfo{person}{Suresh Venkatasubramanian}.}
  \bibinfo{year}{2018}\natexlab{}.
\newblock \showarticletitle{Runaway Feedback Loops in Predictive Policing}. In
  \bibinfo{booktitle}{\emph{Conference of Fairness, Accountability, and
  Transparency}}.
\newblock


\bibitem[\protect\citeauthoryear{Feldman, Friedler, Moeller, Scheidegger, and
  Venkatasubramanian}{Feldman et~al\mbox{.}}{2015}]%
        {feldman2015certifying}
\bibfield{author}{\bibinfo{person}{Michael Feldman}, \bibinfo{person}{Sorelle~A
  Friedler}, \bibinfo{person}{John Moeller}, \bibinfo{person}{Carlos
  Scheidegger}, {and} \bibinfo{person}{Suresh Venkatasubramanian}.}
  \bibinfo{year}{2015}\natexlab{}.
\newblock \showarticletitle{Certifying and removing disparate impact}. In
  \bibinfo{booktitle}{\emph{Proceedings of the 21th ACM SIGKDD International
  Conference on Knowledge Discovery and Data Mining}}.
  \bibinfo{publisher}{ACM}, \bibinfo{pages}{259--268}.
\newblock


\bibitem[\protect\citeauthoryear{Freund and Schapire}{Freund and
  Schapire}{1996}]%
        {freund1996game}
\bibfield{author}{\bibinfo{person}{Yoav Freund} {and} \bibinfo{person}{Robert~E
  Schapire}.} \bibinfo{year}{1996}\natexlab{}.
\newblock \showarticletitle{Game theory, on-line prediction and boosting}. In
  \bibinfo{booktitle}{\emph{COLT}}, Vol.~\bibinfo{volume}{96}.
  \bibinfo{publisher}{Citeseer}, \bibinfo{pages}{325--332}.
\newblock


\bibitem[\protect\citeauthoryear{Friedler, Scheidegger, and
  Venkatasubramanian}{Friedler et~al\mbox{.}}{2016}]%
        {friedler2016possibility}
\bibfield{author}{\bibinfo{person}{Sorelle~A Friedler}, \bibinfo{person}{Carlos
  Scheidegger}, {and} \bibinfo{person}{Suresh Venkatasubramanian}.}
  \bibinfo{year}{2016}\natexlab{}.
\newblock \bibinfo{title}{On the (im) possibility of fairness}.
\newblock
\newblock


\bibitem[\protect\citeauthoryear{Friedler, Scheidegger, Venkatasubramanian,
  Choudhary, Hamilton, and Roth}{Friedler et~al\mbox{.}}{2019}]%
        {friedler2019comparative}
\bibfield{author}{\bibinfo{person}{Sorelle~A Friedler}, \bibinfo{person}{Carlos
  Scheidegger}, \bibinfo{person}{Suresh Venkatasubramanian},
  \bibinfo{person}{Sonam Choudhary}, \bibinfo{person}{Evan~P Hamilton}, {and}
  \bibinfo{person}{Derek Roth}.} \bibinfo{year}{2019}\natexlab{}.
\newblock \showarticletitle{A comparative study of fairness-enhancing
  interventions in machine learning}. In \bibinfo{booktitle}{\emph{Proceedings
  of the Conference on Fairness, Accountability, and Transparency}}.
  \bibinfo{publisher}{ACM}, \bibinfo{pages}{329--338}.
\newblock


\bibitem[\protect\citeauthoryear{Fullinwider}{Fullinwider}{2018}]%
        {sep-affirmative-action}
\bibfield{author}{\bibinfo{person}{Robert Fullinwider}.}
  \bibinfo{year}{2018}\natexlab{}.
\newblock \showarticletitle{Affirmative Action}.
\newblock In \bibinfo{booktitle}{\emph{The Stanford Encyclopedia of Philosophy}
  (\bibinfo{edition}{summer 2018} ed.)},
  \bibfield{editor}{\bibinfo{person}{Edward~N. Zalta}} (Ed.).
  \bibinfo{publisher}{Metaphysics Research Lab, Stanford University}.
\newblock


\bibitem[\protect\citeauthoryear{Goh, Cotter, Gupta, and Friedlander}{Goh
  et~al\mbox{.}}{2016}]%
        {goh2016satisfying}
\bibfield{author}{\bibinfo{person}{Gabriel Goh}, \bibinfo{person}{Andrew
  Cotter}, \bibinfo{person}{Maya Gupta}, {and} \bibinfo{person}{Michael~P
  Friedlander}.} \bibinfo{year}{2016}\natexlab{}.
\newblock \showarticletitle{Satisfying real-world goals with dataset
  constraints}. In \bibinfo{booktitle}{\emph{Advances in Neural Information
  Processing Systems}}. \bibinfo{pages}{2415--2423}.
\newblock


\bibitem[\protect\citeauthoryear{Gonen and Goldberg}{Gonen and
  Goldberg}{2019}]%
        {gonen2019lipstick}
\bibfield{author}{\bibinfo{person}{Hila Gonen} {and} \bibinfo{person}{Yoav
  Goldberg}.} \bibinfo{year}{2019}\natexlab{}.
\newblock \showarticletitle{Lipstick on a Pig: Debiasing Methods Cover up
  Systematic Gender Biases in Word Embeddings But do not Remove Them}. In
  \bibinfo{booktitle}{\emph{Proceedings of the 2019 Conference of the North
  American Chapter of the Association for Computational Linguistics: Human
  Language Technologies, Volume 1 (Long and Short Papers)}}.
  \bibinfo{pages}{609--614}.
\newblock


\bibitem[\protect\citeauthoryear{Goodfellow, Pouget-Abadie, Mirza, Xu,
  Warde-Farley, Ozair, Courville, and Bengio}{Goodfellow et~al\mbox{.}}{2014}]%
        {goodfellow2014generative}
\bibfield{author}{\bibinfo{person}{Ian Goodfellow}, \bibinfo{person}{Jean
  Pouget-Abadie}, \bibinfo{person}{Mehdi Mirza}, \bibinfo{person}{Bing Xu},
  \bibinfo{person}{David Warde-Farley}, \bibinfo{person}{Sherjil Ozair},
  \bibinfo{person}{Aaron Courville}, {and} \bibinfo{person}{Yoshua Bengio}.}
  \bibinfo{year}{2014}\natexlab{}.
\newblock \showarticletitle{Generative adversarial nets}. In
  \bibinfo{booktitle}{\emph{Advances in neural information processing
  systems}}. \bibinfo{pages}{2672--2680}.
\newblock


\bibitem[\protect\citeauthoryear{Gretton, Borgwardt, Rasch, Sch{\"o}lkopf, and
  Smola}{Gretton et~al\mbox{.}}{2012}]%
        {gretton2012kernel}
\bibfield{author}{\bibinfo{person}{Arthur Gretton}, \bibinfo{person}{Karsten~M
  Borgwardt}, \bibinfo{person}{Malte~J Rasch}, \bibinfo{person}{Bernhard
  Sch{\"o}lkopf}, {and} \bibinfo{person}{Alexander Smola}.}
  \bibinfo{year}{2012}\natexlab{}.
\newblock \showarticletitle{A kernel two-sample test}.
\newblock \bibinfo{journal}{\emph{Journal of Machine Learning Research}}
  \bibinfo{volume}{13}, \bibinfo{number}{Mar} (\bibinfo{year}{2012}),
  \bibinfo{pages}{723--773}.
\newblock


\bibitem[\protect\citeauthoryear{Grgic-Hlaca, Redmiles, Gummadi, and
  Weller}{Grgic-Hlaca et~al\mbox{.}}{2018}]%
        {grgic2018human}
\bibfield{author}{\bibinfo{person}{Nina Grgic-Hlaca}, \bibinfo{person}{Elissa~M
  Redmiles}, \bibinfo{person}{Krishna~P Gummadi}, {and} \bibinfo{person}{Adrian
  Weller}.} \bibinfo{year}{2018}\natexlab{}.
\newblock \showarticletitle{Human perceptions of fairness in algorithmic
  decision making: A case study of criminal risk prediction}. In
  \bibinfo{booktitle}{\emph{Proceedings of the 2018 World Wide Web
  Conference}}. \bibinfo{publisher}{International World Wide Web Conferences
  Steering Committee}, \bibinfo{pages}{903--912}.
\newblock


\bibitem[\protect\citeauthoryear{Hamilton}{Hamilton}{2017}]%
        {hamilton2017benchmarking}
\bibfield{author}{\bibinfo{person}{Evan Hamilton}.}
  \bibinfo{year}{2017}\natexlab{}.
\newblock \emph{\bibinfo{title}{Benchmarking Four Approaches to Fairness-Aware
  Machine Learning}}.
\newblock \bibinfo{thesistype}{Ph.D. Dissertation}. \bibinfo{school}{Haverford
  College}.
\newblock


\bibitem[\protect\citeauthoryear{Hardt, Price, and Srebro}{Hardt
  et~al\mbox{.}}{2016}]%
        {hardt2016equality}
\bibfield{author}{\bibinfo{person}{Moritz Hardt}, \bibinfo{person}{Eric Price},
  {and} \bibinfo{person}{Nati Srebro}.} \bibinfo{year}{2016}\natexlab{}.
\newblock \showarticletitle{Equality of opportunity in supervised learning}. In
  \bibinfo{booktitle}{\emph{Advances in neural information processing
  systems}}. \bibinfo{pages}{3315--3323}.
\newblock


\bibitem[\protect\citeauthoryear{Heidari and Krause}{Heidari and
  Krause}{2018}]%
        {heidari2018preventing}
\bibfield{author}{\bibinfo{person}{Hoda Heidari} {and} \bibinfo{person}{Andreas
  Krause}.} \bibinfo{year}{2018}\natexlab{}.
\newblock \showarticletitle{Preventing Disparate Treatment in Sequential
  Decision Making.}. In \bibinfo{booktitle}{\emph{IJCAI}}.
  \bibinfo{pages}{2248--2254}.
\newblock


\bibitem[\protect\citeauthoryear{Hendricks, Burns, Saenko, Darrell, and
  Rohrbach}{Hendricks et~al\mbox{.}}{2018}]%
        {hendricks2018women}
\bibfield{author}{\bibinfo{person}{Lisa~Anne Hendricks},
  \bibinfo{person}{Kaylee Burns}, \bibinfo{person}{Kate Saenko},
  \bibinfo{person}{Trevor Darrell}, {and} \bibinfo{person}{Anna Rohrbach}.}
  \bibinfo{year}{2018}\natexlab{}.
\newblock \showarticletitle{Women also snowboard: Overcoming bias in captioning
  models}. In \bibinfo{booktitle}{\emph{European Conference on Computer
  Vision}}. \bibinfo{publisher}{Springer}, \bibinfo{pages}{793--811}.
\newblock


\bibitem[\protect\citeauthoryear{Holstein, Wortman~Vaughan, Daum{\'e}~III,
  Dudik, and Wallach}{Holstein et~al\mbox{.}}{2019}]%
        {holstein2019improving}
\bibfield{author}{\bibinfo{person}{Kenneth Holstein}, \bibinfo{person}{Jennifer
  Wortman~Vaughan}, \bibinfo{person}{Hal Daum{\'e}~III}, \bibinfo{person}{Miro
  Dudik}, {and} \bibinfo{person}{Hanna Wallach}.}
  \bibinfo{year}{2019}\natexlab{}.
\newblock \showarticletitle{Improving fairness in machine learning systems:
  What do industry practitioners need?}. In
  \bibinfo{booktitle}{\emph{Proceedings of the 2019 CHI Conference on Human
  Factors in Computing Systems}}. \bibinfo{publisher}{ACM},
  \bibinfo{pages}{600}.
\newblock


\bibitem[\protect\citeauthoryear{Hu and Chen}{Hu and Chen}{2018}]%
        {hu2018short}
\bibfield{author}{\bibinfo{person}{Lily Hu} {and} \bibinfo{person}{Yiling
  Chen}.} \bibinfo{year}{2018}\natexlab{}.
\newblock \showarticletitle{A short-term intervention for long-term fairness in
  the labor market}. In \bibinfo{booktitle}{\emph{Proceedings of the 2018 World
  Wide Web Conference}}. \bibinfo{publisher}{International World Wide Web
  Conferences Steering Committee}, \bibinfo{pages}{1389--1398}.
\newblock


\bibitem[\protect\citeauthoryear{Hunt}{Hunt}{2016}]%
        {Hunt:2016:Online}
\bibfield{author}{\bibinfo{person}{Elle Hunt}.}
  \bibinfo{year}{2016}\natexlab{}.
\newblock \bibinfo{title}{Tay, Microsoft's AI chatbot, gets a crash course in
  racism from Twitter}.
\newblock
\newblock
\urldef\tempurl%
\url{https://goo.gl/mE8p3J}
\showURL{%
\tempurl}


\bibitem[\protect\citeauthoryear{Ibarrar{\'a}n, Medell{\'\i}n, Regalia,
  Stampini, Parodi, Tejerina, Cueva, V{\'a}squez, et~al\mbox{.}}{Ibarrar{\'a}n
  et~al\mbox{.}}{2017}]%
        {ibarraran2017conditional}
\bibfield{author}{\bibinfo{person}{Pablo Ibarrar{\'a}n}, \bibinfo{person}{Nadin
  Medell{\'\i}n}, \bibinfo{person}{Ferdinando Regalia}, \bibinfo{person}{Marco
  Stampini}, \bibinfo{person}{Sandro Parodi}, \bibinfo{person}{Luis Tejerina},
  \bibinfo{person}{Pedro Cueva}, \bibinfo{person}{Madiery V{\'a}squez},
  {et~al\mbox{.}}} \bibinfo{year}{2017}\natexlab{}.
\newblock \showarticletitle{How conditional cash transfers work}.
\newblock \bibinfo{journal}{\emph{IDB Publications (Books)}}
  (\bibinfo{year}{2017}).
\newblock


\bibitem[\protect\citeauthoryear{Jabbari, Joseph, Kearns, Morgenstern, and
  Roth}{Jabbari et~al\mbox{.}}{2017}]%
        {jabbari2017fairness}
\bibfield{author}{\bibinfo{person}{Shahin Jabbari}, \bibinfo{person}{Matthew
  Joseph}, \bibinfo{person}{Michael Kearns}, \bibinfo{person}{Jamie
  Morgenstern}, {and} \bibinfo{person}{Aaron Roth}.}
  \bibinfo{year}{2017}\natexlab{}.
\newblock \showarticletitle{Fairness in reinforcement learning}. In
  \bibinfo{booktitle}{\emph{Proceedings of the 34th International Conference on
  Machine Learning-Volume 70}}. \bibinfo{publisher}{JMLR. org},
  \bibinfo{pages}{1617--1626}.
\newblock


\bibitem[\protect\citeauthoryear{Joseph, Kearns, Morgenstern, and Roth}{Joseph
  et~al\mbox{.}}{2016}]%
        {joseph2016fairness}
\bibfield{author}{\bibinfo{person}{Matthew Joseph}, \bibinfo{person}{Michael
  Kearns}, \bibinfo{person}{Jamie~H Morgenstern}, {and} \bibinfo{person}{Aaron
  Roth}.} \bibinfo{year}{2016}\natexlab{}.
\newblock \showarticletitle{Fairness in learning: Classic and contextual
  bandits}. In \bibinfo{booktitle}{\emph{Advances in Neural Information
  Processing Systems}}. \bibinfo{pages}{325--333}.
\newblock


\bibitem[\protect\citeauthoryear{Kallus, Mao, and Zhou}{Kallus
  et~al\mbox{.}}{2019}]%
        {kallus2019assessing}
\bibfield{author}{\bibinfo{person}{Nathan Kallus}, \bibinfo{person}{Xiaojie
  Mao}, {and} \bibinfo{person}{Angela Zhou}.} \bibinfo{year}{2019}\natexlab{}.
\newblock \bibinfo{title}{Assessing Algorithmic Fairness with Unobserved
  Protected Class Using Data Combination}.
\newblock
\newblock


\bibitem[\protect\citeauthoryear{Kamiran and Calders}{Kamiran and
  Calders}{2009}]%
        {kamiran2009classifying}
\bibfield{author}{\bibinfo{person}{Faisal Kamiran} {and} \bibinfo{person}{Toon
  Calders}.} \bibinfo{year}{2009}\natexlab{}.
\newblock \showarticletitle{Classifying without discriminating}. In
  \bibinfo{booktitle}{\emph{2009 2nd International Conference on Computer,
  Control and Communication}}. \bibinfo{publisher}{IEEE},
  \bibinfo{pages}{1--6}.
\newblock


\bibitem[\protect\citeauthoryear{Kamiran and Calders}{Kamiran and
  Calders}{2012}]%
        {kamiran2012data}
\bibfield{author}{\bibinfo{person}{Faisal Kamiran} {and} \bibinfo{person}{Toon
  Calders}.} \bibinfo{year}{2012}\natexlab{}.
\newblock \showarticletitle{Data preprocessing techniques for classification
  without discrimination}.
\newblock \bibinfo{journal}{\emph{Knowledge and Information Systems}}
  \bibinfo{volume}{33}, \bibinfo{number}{1} (\bibinfo{year}{2012}),
  \bibinfo{pages}{1--33}.
\newblock


\bibitem[\protect\citeauthoryear{Kamiran, Calders, and Pechenizkiy}{Kamiran
  et~al\mbox{.}}{2010}]%
        {kamiran2010discrimination}
\bibfield{author}{\bibinfo{person}{Faisal Kamiran}, \bibinfo{person}{Toon
  Calders}, {and} \bibinfo{person}{Mykola Pechenizkiy}.}
  \bibinfo{year}{2010}\natexlab{}.
\newblock \showarticletitle{Discrimination aware decision tree learning}. In
  \bibinfo{booktitle}{\emph{2010 IEEE International Conference on Data
  Mining}}. \bibinfo{publisher}{IEEE}, \bibinfo{pages}{869--874}.
\newblock


\bibitem[\protect\citeauthoryear{Kamishima, Akaho, Asoh, and Sakuma}{Kamishima
  et~al\mbox{.}}{2012a}]%
        {kamishima2012enhancement}
\bibfield{author}{\bibinfo{person}{Toshihiro Kamishima},
  \bibinfo{person}{Shotaro Akaho}, \bibinfo{person}{Hideki Asoh}, {and}
  \bibinfo{person}{Jun Sakuma}.} \bibinfo{year}{2012}\natexlab{a}.
\newblock \showarticletitle{Enhancement of the Neutrality in Recommendation.}.
  In \bibinfo{booktitle}{\emph{Decisions@ RecSys}}. \bibinfo{pages}{8--14}.
\newblock


\bibitem[\protect\citeauthoryear{Kamishima, Akaho, Asoh, and Sakuma}{Kamishima
  et~al\mbox{.}}{2012b}]%
        {kamishima2012fairness}
\bibfield{author}{\bibinfo{person}{Toshihiro Kamishima},
  \bibinfo{person}{Shotaro Akaho}, \bibinfo{person}{Hideki Asoh}, {and}
  \bibinfo{person}{Jun Sakuma}.} \bibinfo{year}{2012}\natexlab{b}.
\newblock \showarticletitle{Fairness-aware classifier with prejudice remover
  regularizer}. In \bibinfo{booktitle}{\emph{Joint European Conference on
  Machine Learning and Knowledge Discovery in Databases}}.
  \bibinfo{publisher}{Springer}, \bibinfo{pages}{35--50}.
\newblock


\bibitem[\protect\citeauthoryear{Kannan, Morgenstern, Roth, Waggoner, and
  Wu}{Kannan et~al\mbox{.}}{2018}]%
        {kannan2018smoothed}
\bibfield{author}{\bibinfo{person}{Sampath Kannan}, \bibinfo{person}{Jamie~H
  Morgenstern}, \bibinfo{person}{Aaron Roth}, \bibinfo{person}{Bo Waggoner},
  {and} \bibinfo{person}{Zhiwei~Steven Wu}.} \bibinfo{year}{2018}\natexlab{}.
\newblock \showarticletitle{A smoothed analysis of the greedy algorithm for the
  linear contextual bandit problem}. In \bibinfo{booktitle}{\emph{Advances in
  Neural Information Processing Systems}}. \bibinfo{pages}{2227--2236}.
\newblock


\bibitem[\protect\citeauthoryear{K{\"a}rkk{\"a}inen and Joo}{K{\"a}rkk{\"a}inen
  and Joo}{2019}]%
        {karkkainen2019fairface}
\bibfield{author}{\bibinfo{person}{Kimmo K{\"a}rkk{\"a}inen} {and}
  \bibinfo{person}{Jungseock Joo}.} \bibinfo{year}{2019}\natexlab{}.
\newblock \bibinfo{title}{FairFace: Face Attribute Dataset for Balanced Race,
  Gender, and Age}.
\newblock
\newblock


\bibitem[\protect\citeauthoryear{Kay, Matuszek, and Munson}{Kay
  et~al\mbox{.}}{2015}]%
        {kay2015unequal}
\bibfield{author}{\bibinfo{person}{Matthew Kay}, \bibinfo{person}{Cynthia
  Matuszek}, {and} \bibinfo{person}{Sean~A Munson}.}
  \bibinfo{year}{2015}\natexlab{}.
\newblock \showarticletitle{Unequal representation and gender stereotypes in
  image search results for occupations}. In
  \bibinfo{booktitle}{\emph{Proceedings of the 33rd Annual ACM Conference on
  Human Factors in Computing Systems}}. \bibinfo{publisher}{ACM},
  \bibinfo{pages}{3819--3828}.
\newblock


\bibitem[\protect\citeauthoryear{Kazemi, Zadimoghaddam, and Karbasi}{Kazemi
  et~al\mbox{.}}{2018}]%
        {kazemi2018scalable}
\bibfield{author}{\bibinfo{person}{Ehsan Kazemi}, \bibinfo{person}{Morteza
  Zadimoghaddam}, {and} \bibinfo{person}{Amin Karbasi}.}
  \bibinfo{year}{2018}\natexlab{}.
\newblock \showarticletitle{Scalable deletion-robust submodular maximization:
  Data summarization with privacy and fairness constraints}. In
  \bibinfo{booktitle}{\emph{International conference on machine learning}}.
  \bibinfo{pages}{2549--2558}.
\newblock


\bibitem[\protect\citeauthoryear{Kilbertus, Carulla, Parascandolo, Hardt,
  Janzing, and Sch{\"o}lkopf}{Kilbertus et~al\mbox{.}}{2017}]%
        {kilbertus2017avoiding}
\bibfield{author}{\bibinfo{person}{Niki Kilbertus},
  \bibinfo{person}{Mateo~Rojas Carulla}, \bibinfo{person}{Giambattista
  Parascandolo}, \bibinfo{person}{Moritz Hardt}, \bibinfo{person}{Dominik
  Janzing}, {and} \bibinfo{person}{Bernhard Sch{\"o}lkopf}.}
  \bibinfo{year}{2017}\natexlab{}.
\newblock \showarticletitle{Avoiding discrimination through causal reasoning}.
  In \bibinfo{booktitle}{\emph{Advances in Neural Information Processing
  Systems}}. \bibinfo{pages}{656--666}.
\newblock


\bibitem[\protect\citeauthoryear{Kleinberg, Mullainathan, and
  Raghavan}{Kleinberg et~al\mbox{.}}{2017}]%
        {kleinberg2017inherent}
\bibfield{author}{\bibinfo{person}{Jon Kleinberg}, \bibinfo{person}{Sendhil
  Mullainathan}, {and} \bibinfo{person}{Manish Raghavan}.}
  \bibinfo{year}{2017}\natexlab{}.
\newblock \showarticletitle{Inherent Trade-Offs in the Fair Determination of
  Risk Scores}. In \bibinfo{booktitle}{\emph{8th Innovations in Theoretical
  Computer Science Conference (ITCS 2017)}}. \bibinfo{publisher}{Schloss
  Dagstuhl-Leibniz-Zentrum fuer Informatik}.
\newblock


\bibitem[\protect\citeauthoryear{Kunaver and Po{\v{z}}rl}{Kunaver and
  Po{\v{z}}rl}{2017}]%
        {kunaver2017diversity}
\bibfield{author}{\bibinfo{person}{Matev{\v{z}} Kunaver} {and}
  \bibinfo{person}{Toma{\v{z}} Po{\v{z}}rl}.} \bibinfo{year}{2017}\natexlab{}.
\newblock \showarticletitle{Diversity in recommender systems--A survey}.
\newblock \bibinfo{journal}{\emph{Knowledge-Based Systems}}
  \bibinfo{volume}{123} (\bibinfo{year}{2017}), \bibinfo{pages}{154--162}.
\newblock


\bibitem[\protect\citeauthoryear{Kusner, Loftus, Russell, and Silva}{Kusner
  et~al\mbox{.}}{2017}]%
        {kusner2017counterfactual}
\bibfield{author}{\bibinfo{person}{Matt~J Kusner}, \bibinfo{person}{Joshua
  Loftus}, \bibinfo{person}{Chris Russell}, {and} \bibinfo{person}{Ricardo
  Silva}.} \bibinfo{year}{2017}\natexlab{}.
\newblock \showarticletitle{Counterfactual fairness}. In
  \bibinfo{booktitle}{\emph{Advances in Neural Information Processing
  Systems}}. \bibinfo{pages}{4066--4076}.
\newblock


\bibitem[\protect\citeauthoryear{Larson, Mattu, Kirchner, and Angwin}{Larson
  et~al\mbox{.}}{2016}]%
        {Larson:2016:Online}
\bibfield{author}{\bibinfo{person}{Jeff Larson}, \bibinfo{person}{Surya Mattu},
  \bibinfo{person}{Lauren Kirchner}, {and} \bibinfo{person}{Julia Angwin}.}
  \bibinfo{year}{2016}\natexlab{}.
\newblock \bibinfo{title}{How We Analyzed the COMPAS Recidivism Algorithm}.
\newblock
\newblock
\urldef\tempurl%
\url{https://www.propublica.org/article/how-we-analyzed-the-compas-recidivism-algorithm}
\showURL{%
\tempurl}


\bibitem[\protect\citeauthoryear{Lee, Lou, Chen, Chen, Lin, Chiang, and
  Chen}{Lee et~al\mbox{.}}{2014}]%
        {lee2014fairness}
\bibfield{author}{\bibinfo{person}{Eric~L Lee}, \bibinfo{person}{Jing-Kai Lou},
  \bibinfo{person}{Wei-Ming Chen}, \bibinfo{person}{Yen-Chi Chen},
  \bibinfo{person}{Shou-De Lin}, \bibinfo{person}{Yen-Sheng Chiang}, {and}
  \bibinfo{person}{Kuan-Ta Chen}.} \bibinfo{year}{2014}\natexlab{}.
\newblock \showarticletitle{Fairness-aware loan recommendation for microfinance
  services}. In \bibinfo{booktitle}{\emph{Proceedings of the 2014 International
  Conference on Social Computing}}. \bibinfo{publisher}{ACM},
  \bibinfo{pages}{3}.
\newblock


\bibitem[\protect\citeauthoryear{Lehmann and Romano}{Lehmann and
  Romano}{2006}]%
        {lehmann2006testing}
\bibfield{author}{\bibinfo{person}{Erich~L Lehmann} {and}
  \bibinfo{person}{Joseph~P Romano}.} \bibinfo{year}{2006}\natexlab{}.
\newblock \bibinfo{booktitle}{\emph{Testing statistical hypotheses}}.
\newblock \bibinfo{publisher}{Springer Science \& Business Media}.
\newblock


\bibitem[\protect\citeauthoryear{Lipton, Chouldechova, and McAuley}{Lipton
  et~al\mbox{.}}{2017}]%
        {lipton2017does}
\bibfield{author}{\bibinfo{person}{Zachary~C Lipton},
  \bibinfo{person}{Alexandra Chouldechova}, {and} \bibinfo{person}{Julian
  McAuley}.} \bibinfo{year}{2017}\natexlab{}.
\newblock \showarticletitle{Does mitigating ML’s disparate impact require
  disparate treatment?}
\newblock \bibinfo{journal}{\emph{stat}}  \bibinfo{volume}{1050}
  (\bibinfo{year}{2017}), \bibinfo{pages}{19}.
\newblock


\bibitem[\protect\citeauthoryear{Liu and Burke}{Liu and Burke}{2018}]%
        {liu2018personalizing}
\bibfield{author}{\bibinfo{person}{Weiwen Liu} {and} \bibinfo{person}{Robin
  Burke}.} \bibinfo{year}{2018}\natexlab{}.
\newblock \bibinfo{title}{Personalizing fairness-aware re-ranking}.
\newblock
\newblock


\bibitem[\protect\citeauthoryear{Loftus, Russell, Kusner, and Silva}{Loftus
  et~al\mbox{.}}{2018}]%
        {loftus2018causal}
\bibfield{author}{\bibinfo{person}{Joshua~R Loftus}, \bibinfo{person}{Chris
  Russell}, \bibinfo{person}{Matt~J Kusner}, {and} \bibinfo{person}{Ricardo
  Silva}.} \bibinfo{year}{2018}\natexlab{}.
\newblock \bibinfo{title}{Causal reasoning for algorithmic fairness}.
\newblock
\newblock


\bibitem[\protect\citeauthoryear{Louizos, Swersky, Li, Welling, and
  Zemel}{Louizos et~al\mbox{.}}{2016}]%
        {louizos2016variational}
\bibfield{author}{\bibinfo{person}{Christos Louizos}, \bibinfo{person}{Kevin
  Swersky}, \bibinfo{person}{Yujia Li}, \bibinfo{person}{Max Welling}, {and}
  \bibinfo{person}{Richard Zemel}.} \bibinfo{year}{2016}\natexlab{}.
\newblock \showarticletitle{The variational fair autoencoder}.
\newblock \bibinfo{journal}{\emph{International Conference on Learning
  Representations (ICLR)}} (\bibinfo{year}{2016}).
\newblock


\bibitem[\protect\citeauthoryear{Luong, Ruggieri, and Turini}{Luong
  et~al\mbox{.}}{2011}]%
        {luong2011k}
\bibfield{author}{\bibinfo{person}{Binh~Thanh Luong},
  \bibinfo{person}{Salvatore Ruggieri}, {and} \bibinfo{person}{Franco Turini}.}
  \bibinfo{year}{2011}\natexlab{}.
\newblock \showarticletitle{k-NN as an implementation of situation testing for
  discrimination discovery and prevention}. In
  \bibinfo{booktitle}{\emph{Proceedings of the 17th ACM SIGKDD international
  conference on Knowledge discovery and data mining}}.
  \bibinfo{publisher}{ACM}, \bibinfo{pages}{502--510}.
\newblock


\bibitem[\protect\citeauthoryear{Madras, Creager, Pitassi, and Zemel}{Madras
  et~al\mbox{.}}{2018a}]%
        {madras2018learning}
\bibfield{author}{\bibinfo{person}{David Madras}, \bibinfo{person}{Elliot
  Creager}, \bibinfo{person}{Toniann Pitassi}, {and} \bibinfo{person}{Richard
  Zemel}.} \bibinfo{year}{2018}\natexlab{a}.
\newblock \bibinfo{title}{Learning adversarially fair and transferable
  representations}.
\newblock
\newblock


\bibitem[\protect\citeauthoryear{Madras, Pitassi, and Zemel}{Madras
  et~al\mbox{.}}{2018b}]%
        {madras2018predict}
\bibfield{author}{\bibinfo{person}{David Madras}, \bibinfo{person}{Toniann
  Pitassi}, {and} \bibinfo{person}{Richard Zemel}.}
  \bibinfo{year}{2018}\natexlab{b}.
\newblock \showarticletitle{Predict responsibly: Increasing fairness by
  learning to defer}.
\newblock \bibinfo{journal}{\emph{International Conference on Learning
  Representations (ICLR)}} (\bibinfo{year}{2018}).
\newblock


\bibitem[\protect\citeauthoryear{Manthorpe}{Manthorpe}{2017}]%
        {Manthorpe:2017:Online}
\bibfield{author}{\bibinfo{person}{Rowland Manthorpe}.}
  \bibinfo{year}{2017}\natexlab{}.
\newblock \bibinfo{title}{Beauty.AI's 'robot beauty contest' is back – and
  this time it promises not to be racist}.
\newblock
\newblock
\urldef\tempurl%
\url{https://www.wired.co.uk/article/robot-beauty-contest-beauty-ai}
\showURL{%
\tempurl}


\bibitem[\protect\citeauthoryear{Mart{\'\i}nez-Plumed, Ferri, Nieves, and
  Hern{\'a}ndez-Orallo}{Mart{\'\i}nez-Plumed et~al\mbox{.}}{2019}]%
        {martinez2019fairness}
\bibfield{author}{\bibinfo{person}{Fernando Mart{\'\i}nez-Plumed},
  \bibinfo{person}{C{\`e}sar Ferri}, \bibinfo{person}{David Nieves}, {and}
  \bibinfo{person}{Jos{\'e} Hern{\'a}ndez-Orallo}.}
  \bibinfo{year}{2019}\natexlab{}.
\newblock \bibinfo{title}{Fairness and Missing Values}.
\newblock
\newblock


\bibitem[\protect\citeauthoryear{May, Wang, Bordia, Bowman, and Rudinger}{May
  et~al\mbox{.}}{2019}]%
        {may2019measuring}
\bibfield{author}{\bibinfo{person}{Chandler May}, \bibinfo{person}{Alex Wang},
  \bibinfo{person}{Shikha Bordia}, \bibinfo{person}{Samuel Bowman}, {and}
  \bibinfo{person}{Rachel Rudinger}.} \bibinfo{year}{2019}\natexlab{}.
\newblock \showarticletitle{On Measuring Social Biases in Sentence Encoders}.
  In \bibinfo{booktitle}{\emph{Proceedings of the 2019 Conference of the North
  American Chapter of the Association for Computational Linguistics: Human
  Language Technologies, Volume 1 (Long and Short Papers)}}.
  \bibinfo{pages}{622--628}.
\newblock


\bibitem[\protect\citeauthoryear{Menon and Williamson}{Menon and
  Williamson}{2018}]%
        {menon2018cost}
\bibfield{author}{\bibinfo{person}{Aditya~Krishna Menon} {and}
  \bibinfo{person}{Robert~C Williamson}.} \bibinfo{year}{2018}\natexlab{}.
\newblock \showarticletitle{The cost of fairness in binary classification}. In
  \bibinfo{booktitle}{\emph{Conference on Fairness, Accountability and
  Transparency}}. \bibinfo{pages}{107--118}.
\newblock


\bibitem[\protect\citeauthoryear{Miao}{Miao}{2011}]%
        {miao2011did}
\bibfield{author}{\bibinfo{person}{Weiwen Miao}.}
  \bibinfo{year}{2011}\natexlab{}.
\newblock \showarticletitle{Did the results of promotion exams have a disparate
  impact on minorities? Using statistical evidence in Ricci v}.
\newblock \bibinfo{journal}{\emph{DeStefano. J. of Stat. Ed}}
  \bibinfo{volume}{19}, \bibinfo{number}{1} (\bibinfo{year}{2011}).
\newblock


\bibitem[\protect\citeauthoryear{Moro, Cortez, and Rita}{Moro
  et~al\mbox{.}}{2014}]%
        {moro2014data}
\bibfield{author}{\bibinfo{person}{S{\'e}rgio Moro}, \bibinfo{person}{Paulo
  Cortez}, {and} \bibinfo{person}{Paulo Rita}.}
  \bibinfo{year}{2014}\natexlab{}.
\newblock \showarticletitle{A data-driven approach to predict the success of
  bank telemarketing}.
\newblock \bibinfo{journal}{\emph{Decision Support Systems}}
  \bibinfo{volume}{62} (\bibinfo{year}{2014}), \bibinfo{pages}{22--31}.
\newblock


\bibitem[\protect\citeauthoryear{Mullainathan}{Mullainathan}{2019}]%
        {Mullainathan:2019:Online}
\bibfield{author}{\bibinfo{person}{Sendhil Mullainathan}.}
  \bibinfo{year}{2019}\natexlab{}.
\newblock \bibinfo{title}{Biased Algorithms Are Easier to Fix Than Biased
  People}.
\newblock
\newblock
\urldef\tempurl%
\url{https://www.nytimes.com/2019/12/06/business/algorithm-bias-fix.html?smid=nytcore-ios-share}
\showURL{%
\tempurl}


\bibitem[\protect\citeauthoryear{Nabi and Shpitser}{Nabi and Shpitser}{2018}]%
        {nabi2018fair}
\bibfield{author}{\bibinfo{person}{Razieh Nabi} {and} \bibinfo{person}{Ilya
  Shpitser}.} \bibinfo{year}{2018}\natexlab{}.
\newblock \showarticletitle{Fair inference on outcomes}. In
  \bibinfo{booktitle}{\emph{Thirty-Second AAAI Conference on Artificial
  Intelligence}}.
\newblock


\bibitem[\protect\citeauthoryear{Noriega-Campero, Bakker, Garcia-Bulle, and
  Pentland}{Noriega-Campero et~al\mbox{.}}{2019}]%
        {noriega2019active}
\bibfield{author}{\bibinfo{person}{Alejandro Noriega-Campero},
  \bibinfo{person}{Michiel~A Bakker}, \bibinfo{person}{Bernardo Garcia-Bulle},
  {and} \bibinfo{person}{Alex'Sandy' Pentland}.}
  \bibinfo{year}{2019}\natexlab{}.
\newblock \showarticletitle{Active Fairness in Algorithmic Decision Making}. In
  \bibinfo{booktitle}{\emph{Proceedings of the 2019 AAAI/ACM Conference on AI,
  Ethics, and Society}}. \bibinfo{publisher}{ACM}, \bibinfo{pages}{77--83}.
\newblock


\bibitem[\protect\citeauthoryear{Pachal}{Pachal}{2015}]%
        {Pachal:2015:Online}
\bibfield{author}{\bibinfo{person}{Pete Pachal}.}
  \bibinfo{year}{2015}\natexlab{}.
\newblock \bibinfo{title}{Google Photos identified two black people as
  'gorillas'}.
\newblock
\newblock
\urldef\tempurl%
\url{https://mashable.com/2015/07/01/google-photos-black-people-gorillas/}
\showURL{%
\tempurl}


\bibitem[\protect\citeauthoryear{Pan and Yang}{Pan and Yang}{2009}]%
        {pan2009survey}
\bibfield{author}{\bibinfo{person}{Sinno~Jialin Pan} {and}
  \bibinfo{person}{Qiang Yang}.} \bibinfo{year}{2009}\natexlab{}.
\newblock \showarticletitle{A survey on transfer learning}.
\newblock \bibinfo{journal}{\emph{IEEE Transactions on knowledge and data
  engineering}} \bibinfo{volume}{22}, \bibinfo{number}{10}
  (\bibinfo{year}{2009}), \bibinfo{pages}{1345--1359}.
\newblock


\bibitem[\protect\citeauthoryear{Pennington, Socher, and Manning}{Pennington
  et~al\mbox{.}}{2014}]%
        {pennington2014glove}
\bibfield{author}{\bibinfo{person}{Jeffrey Pennington},
  \bibinfo{person}{Richard Socher}, {and} \bibinfo{person}{Christopher
  Manning}.} \bibinfo{year}{2014}\natexlab{}.
\newblock \showarticletitle{Glove: Global vectors for word representation}. In
  \bibinfo{booktitle}{\emph{Proceedings of the 2014 conference on empirical
  methods in natural language processing (EMNLP)}}.
  \bibinfo{pages}{1532--1543}.
\newblock


\bibitem[\protect\citeauthoryear{Pessach and Shmueli}{Pessach and
  Shmueli}{2020}]%
        {Pessach2020fairselection}
\bibfield{author}{\bibinfo{person}{D Pessach} {and} \bibinfo{person}{E
  Shmueli}.} \bibinfo{year}{2020}\natexlab{}.
\newblock \bibinfo{title}{Improving Fairness in Semi-Supervised Problems with
  Privileged-Group Selection Bias}.  (\bibinfo{year}{2020}).
\newblock
\newblock
\shownote{Working paper.}


\bibitem[\protect\citeauthoryear{Plaugic}{Plaugic}{2017}]%
        {Plaugic:2017:Online}
\bibfield{author}{\bibinfo{person}{Lizzie Plaugic}.}
  \bibinfo{year}{2017}\natexlab{}.
\newblock \bibinfo{title}{FaceApp's creator apologizes for the app's
  skin-lightening 'hot' filter}.
\newblock
\newblock
\urldef\tempurl%
\url{https://www.theverge.com/2017/4/25/15419522/faceapp-hot-filter-racist-apology}
\showURL{%
\tempurl}


\bibitem[\protect\citeauthoryear{Pleiss, Raghavan, Wu, Kleinberg, and
  Weinberger}{Pleiss et~al\mbox{.}}{2017}]%
        {pleiss2017fairness}
\bibfield{author}{\bibinfo{person}{Geoff Pleiss}, \bibinfo{person}{Manish
  Raghavan}, \bibinfo{person}{Felix Wu}, \bibinfo{person}{Jon Kleinberg}, {and}
  \bibinfo{person}{Kilian~Q Weinberger}.} \bibinfo{year}{2017}\natexlab{}.
\newblock \showarticletitle{On fairness and calibration}. In
  \bibinfo{booktitle}{\emph{Advances in Neural Information Processing
  Systems}}. \bibinfo{pages}{5680--5689}.
\newblock


\bibitem[\protect\citeauthoryear{Quadrianto and Sharmanska}{Quadrianto and
  Sharmanska}{2017}]%
        {quadrianto2017recycling}
\bibfield{author}{\bibinfo{person}{Novi Quadrianto} {and}
  \bibinfo{person}{Viktoriia Sharmanska}.} \bibinfo{year}{2017}\natexlab{}.
\newblock \showarticletitle{Recycling privileged learning and distribution
  matching for fairness}. In \bibinfo{booktitle}{\emph{Advances in Neural
  Information Processing Systems}}. \bibinfo{pages}{677--688}.
\newblock


\bibitem[\protect\citeauthoryear{Quadrianto, Sharmanska, and Thomas}{Quadrianto
  et~al\mbox{.}}{2019}]%
        {quadrianto2019discovering}
\bibfield{author}{\bibinfo{person}{Novi Quadrianto}, \bibinfo{person}{Viktoriia
  Sharmanska}, {and} \bibinfo{person}{Oliver Thomas}.}
  \bibinfo{year}{2019}\natexlab{}.
\newblock \showarticletitle{Discovering fair representations in the data
  domain}. In \bibinfo{booktitle}{\emph{Proceedings of the IEEE Conference on
  Computer Vision and Pattern Recognition}}. \bibinfo{pages}{8227--8236}.
\newblock


\bibitem[\protect\citeauthoryear{Redmond and Baveja}{Redmond and
  Baveja}{2002}]%
        {redmond2002data}
\bibfield{author}{\bibinfo{person}{Michael Redmond} {and} \bibinfo{person}{Alok
  Baveja}.} \bibinfo{year}{2002}\natexlab{}.
\newblock \showarticletitle{A data-driven software tool for enabling
  cooperative information sharing among police departments}.
\newblock \bibinfo{journal}{\emph{European Journal of Operational Research}}
  \bibinfo{volume}{141}, \bibinfo{number}{3} (\bibinfo{year}{2002}),
  \bibinfo{pages}{660--678}.
\newblock


\bibitem[\protect\citeauthoryear{Roth}{Roth}{2018}]%
        {roth2018comparison}
\bibfield{author}{\bibinfo{person}{Derek Roth}.}
  \bibinfo{year}{2018}\natexlab{}.
\newblock \emph{\bibinfo{title}{A Comparison of Fairness-Aware Machine Learning
  Algorithms}}.
\newblock \bibinfo{thesistype}{Ph.D. Dissertation}. \bibinfo{school}{Haverford
  College}.
\newblock


\bibitem[\protect\citeauthoryear{Rudinger, Naradowsky, Leonard, and
  Van~Durme}{Rudinger et~al\mbox{.}}{2018}]%
        {rudinger2018gender}
\bibfield{author}{\bibinfo{person}{Rachel Rudinger}, \bibinfo{person}{Jason
  Naradowsky}, \bibinfo{person}{Brian Leonard}, {and} \bibinfo{person}{Benjamin
  Van~Durme}.} \bibinfo{year}{2018}\natexlab{}.
\newblock \showarticletitle{Gender Bias in Coreference Resolution}. In
  \bibinfo{booktitle}{\emph{Proceedings of the 2018 Conference of the North
  American Chapter of the Association for Computational Linguistics: Human
  Language Technologies, Volume 2 (Short Papers)}}. \bibinfo{pages}{8--14}.
\newblock


\bibitem[\protect\citeauthoryear{Russell, Kusner, Loftus, and Silva}{Russell
  et~al\mbox{.}}{2017}]%
        {russell2017worlds}
\bibfield{author}{\bibinfo{person}{Chris Russell}, \bibinfo{person}{Matt~J
  Kusner}, \bibinfo{person}{Joshua Loftus}, {and} \bibinfo{person}{Ricardo
  Silva}.} \bibinfo{year}{2017}\natexlab{}.
\newblock \showarticletitle{When worlds collide: integrating different
  counterfactual assumptions in fairness}. In
  \bibinfo{booktitle}{\emph{Advances in Neural Information Processing
  Systems}}. \bibinfo{pages}{6414--6423}.
\newblock


\bibitem[\protect\citeauthoryear{Rutherglen}{Rutherglen}{1987}]%
        {rutherglen1987disparate}
\bibfield{author}{\bibinfo{person}{George Rutherglen}.}
  \bibinfo{year}{1987}\natexlab{}.
\newblock \showarticletitle{Disparate impact under title VII: an objective
  theory of discrimination}.
\newblock \bibinfo{journal}{\emph{Va. L. Rev.}}  \bibinfo{volume}{73}
  (\bibinfo{year}{1987}), \bibinfo{pages}{1297}.
\newblock


\bibitem[\protect\citeauthoryear{Rutherglen}{Rutherglen}{2009}]%
        {rutherglen2009ricci}
\bibfield{author}{\bibinfo{person}{George Rutherglen}.}
  \bibinfo{year}{2009}\natexlab{}.
\newblock \showarticletitle{Ricci v DeStefano: Affirmative Action and the
  Lessons of Adversity}.
\newblock \bibinfo{journal}{\emph{The Supreme Court Review}}
  \bibinfo{volume}{2009}, \bibinfo{number}{1} (\bibinfo{year}{2009}),
  \bibinfo{pages}{83--114}.
\newblock


\bibitem[\protect\citeauthoryear{Ryu, Adam, and Mitchell}{Ryu
  et~al\mbox{.}}{2017}]%
        {ryu2017inclusivefacenet}
\bibfield{author}{\bibinfo{person}{Hee~Jung Ryu}, \bibinfo{person}{Hartwig
  Adam}, {and} \bibinfo{person}{Margaret Mitchell}.}
  \bibinfo{year}{2017}\natexlab{}.
\newblock \bibinfo{title}{Inclusivefacenet: Improving face attribute detection
  with race and gender diversity}.
\newblock
\newblock


\bibitem[\protect\citeauthoryear{Samadi, Tantipongpipat, Morgenstern, Singh,
  and Vempala}{Samadi et~al\mbox{.}}{2018}]%
        {samadi2018price}
\bibfield{author}{\bibinfo{person}{Samira Samadi}, \bibinfo{person}{Uthaipon
  Tantipongpipat}, \bibinfo{person}{Jamie~H Morgenstern},
  \bibinfo{person}{Mohit Singh}, {and} \bibinfo{person}{Santosh Vempala}.}
  \bibinfo{year}{2018}\natexlab{}.
\newblock \showarticletitle{The price of fair PCA: One extra dimension}. In
  \bibinfo{booktitle}{\emph{Advances in Neural Information Processing
  Systems}}. \bibinfo{pages}{10976--10987}.
\newblock


\bibitem[\protect\citeauthoryear{Sander}{Sander}{2004}]%
        {sander2004systemic}
\bibfield{author}{\bibinfo{person}{Richard~H Sander}.}
  \bibinfo{year}{2004}\natexlab{}.
\newblock \showarticletitle{A systemic analysis of affirmative action in
  American law schools}.
\newblock \bibinfo{journal}{\emph{Stan. L. Rev.}}  \bibinfo{volume}{57}
  (\bibinfo{year}{2004}), \bibinfo{pages}{367}.
\newblock


\bibitem[\protect\citeauthoryear{Sattigeri, Hoffman, Chenthamarakshan, and
  Varshney}{Sattigeri et~al\mbox{.}}{2019}]%
        {sattigeri2019fairness}
\bibfield{author}{\bibinfo{person}{Prasanna Sattigeri},
  \bibinfo{person}{Samuel~C Hoffman}, \bibinfo{person}{Vijil Chenthamarakshan},
  {and} \bibinfo{person}{Kush~Raj Varshney}.} \bibinfo{year}{2019}\natexlab{}.
\newblock \showarticletitle{Fairness GAN: Generating datasets with fairness
  properties using a generative adversarial network}.
\newblock \bibinfo{journal}{\emph{IBM Journal of Research and Development}}
  (\bibinfo{year}{2019}).
\newblock


\bibitem[\protect\citeauthoryear{Simonite}{Simonite}{2015}]%
        {Simonite:2015:Online}
\bibfield{author}{\bibinfo{person}{Tom Simonite}.}
  \bibinfo{year}{2015}\natexlab{}.
\newblock \bibinfo{title}{Probing the Dark Side of Google’s Ad-Targeting
  System}.
\newblock
\newblock
\urldef\tempurl%
\url{https://www.technologyreview.com/s/539021/probing-the-dark-side-of-googles-ad-targeting-system/}
\showURL{%
\tempurl}


\bibitem[\protect\citeauthoryear{Simonite}{Simonite}{2018}]%
        {Simonite:2018:Online}
\bibfield{author}{\bibinfo{person}{Tom Simonite}.}
  \bibinfo{year}{2018}\natexlab{}.
\newblock \bibinfo{title}{When It Comes to Gorillas, Google Photos Remains
  Blind}.
\newblock
\newblock
\urldef\tempurl%
\url{https://www.wired.com/story/when-it-comes-to-gorillas-google-photos-remains-blind/}
\showURL{%
\tempurl}


\bibitem[\protect\citeauthoryear{Spirtes, Meek, and Richardson}{Spirtes
  et~al\mbox{.}}{1995}]%
        {spirtes1995causal}
\bibfield{author}{\bibinfo{person}{Peter Spirtes}, \bibinfo{person}{Christopher
  Meek}, {and} \bibinfo{person}{Thomas Richardson}.}
  \bibinfo{year}{1995}\natexlab{}.
\newblock \showarticletitle{Causal inference in the presence of latent
  variables and selection bias}. In \bibinfo{booktitle}{\emph{Proceedings of
  the Eleventh conference on Uncertainty in artificial intelligence}}.
  \bibinfo{publisher}{Morgan Kaufmann Publishers Inc.},
  \bibinfo{pages}{499--506}.
\newblock


\bibitem[\protect\citeauthoryear{Srivastava, Heidari, and Krause}{Srivastava
  et~al\mbox{.}}{2019}]%
        {srivastava2019mathematical}
\bibfield{author}{\bibinfo{person}{Megha Srivastava}, \bibinfo{person}{Hoda
  Heidari}, {and} \bibinfo{person}{Andreas Krause}.}
  \bibinfo{year}{2019}\natexlab{}.
\newblock \bibinfo{title}{Mathematical Notions vs. Human Perception of
  Fairness: A Descriptive Approach to Fairness for Machine Learning}.
\newblock
\newblock


\bibitem[\protect\citeauthoryear{Stock and Cisse}{Stock and Cisse}{2017}]%
        {stock2017convnets}
\bibfield{author}{\bibinfo{person}{Pierre Stock} {and}
  \bibinfo{person}{Moustapha Cisse}.} \bibinfo{year}{2017}\natexlab{}.
\newblock \bibinfo{title}{Convnets and imagenet beyond accuracy: Explanations,
  bias detection, adversarial examples and model criticism}.
\newblock
\newblock


\bibitem[\protect\citeauthoryear{S{\"u}rer, Burke, and Malthouse}{S{\"u}rer
  et~al\mbox{.}}{2018}]%
        {surer2018multistakeholder}
\bibfield{author}{\bibinfo{person}{{\"O}zge S{\"u}rer}, \bibinfo{person}{Robin
  Burke}, {and} \bibinfo{person}{Edward~C Malthouse}.}
  \bibinfo{year}{2018}\natexlab{}.
\newblock \showarticletitle{Multistakeholder recommendation with provider
  constraints}. In \bibinfo{booktitle}{\emph{Proceedings of the 12th ACM
  Conference on Recommender Systems}}. \bibinfo{publisher}{ACM},
  \bibinfo{pages}{54--62}.
\newblock


\bibitem[\protect\citeauthoryear{TISHBY}{TISHBY}{1999}]%
        {tishby1999information}
\bibfield{author}{\bibinfo{person}{N TISHBY}.} \bibinfo{year}{1999}\natexlab{}.
\newblock \showarticletitle{The information bottleneck method}. In
  \bibinfo{booktitle}{\emph{Proc. 37th Annual Allerton Conference on
  Communications, Control and Computing, 1999}}. \bibinfo{pages}{368--377}.
\newblock


\bibitem[\protect\citeauthoryear{Tramer, Atlidakis, Geambasu, Hsu, Hubaux,
  Humbert, Juels, and Lin}{Tramer et~al\mbox{.}}{2017}]%
        {tramer2017fairtest}
\bibfield{author}{\bibinfo{person}{Florian Tramer}, \bibinfo{person}{Vaggelis
  Atlidakis}, \bibinfo{person}{Roxana Geambasu}, \bibinfo{person}{Daniel Hsu},
  \bibinfo{person}{Jean-Pierre Hubaux}, \bibinfo{person}{Mathias Humbert},
  \bibinfo{person}{Ari Juels}, {and} \bibinfo{person}{Huang Lin}.}
  \bibinfo{year}{2017}\natexlab{}.
\newblock \showarticletitle{FairTest: Discovering unwarranted associations in
  data-driven applications}. In \bibinfo{booktitle}{\emph{2017 IEEE European
  Symposium on Security and Privacy (EuroS\&P)}}. \bibinfo{publisher}{IEEE},
  \bibinfo{pages}{401--416}.
\newblock


\bibitem[\protect\citeauthoryear{Valera, Singla, and Rodriguez}{Valera
  et~al\mbox{.}}{2018}]%
        {valera2018enhancing}
\bibfield{author}{\bibinfo{person}{Isabel Valera}, \bibinfo{person}{Adish
  Singla}, {and} \bibinfo{person}{Manuel~Gomez Rodriguez}.}
  \bibinfo{year}{2018}\natexlab{}.
\newblock \showarticletitle{Enhancing the accuracy and fairness of human
  decision making}. In \bibinfo{booktitle}{\emph{Advances in Neural Information
  Processing Systems}}. \bibinfo{pages}{1769--1778}.
\newblock


\bibitem[\protect\citeauthoryear{van Miltenburg}{van Miltenburg}{2016}]%
        {van2016stereotyping}
\bibfield{author}{\bibinfo{person}{Emiel van Miltenburg}.}
  \bibinfo{year}{2016}\natexlab{}.
\newblock \bibinfo{title}{Stereotyping and Bias in the Flickr30k Dataset}.
\newblock , \bibinfo{numpages}{4}~pages.
\newblock


\bibitem[\protect\citeauthoryear{Vapnik and Izmailov}{Vapnik and
  Izmailov}{2015}]%
        {vapnik2015learning}
\bibfield{author}{\bibinfo{person}{Vladimir Vapnik} {and} \bibinfo{person}{Rauf
  Izmailov}.} \bibinfo{year}{2015}\natexlab{}.
\newblock \showarticletitle{Learning using privileged information: similarity
  control and knowledge transfer.}
\newblock \bibinfo{journal}{\emph{Journal of machine learning research}}
  \bibinfo{volume}{16}, \bibinfo{number}{2023-2049} (\bibinfo{year}{2015}),
  \bibinfo{pages}{2}.
\newblock


\bibitem[\protect\citeauthoryear{Verma and Rubin}{Verma and Rubin}{2018}]%
        {verma2018fairness}
\bibfield{author}{\bibinfo{person}{Sahil Verma} {and} \bibinfo{person}{Julia
  Rubin}.} \bibinfo{year}{2018}\natexlab{}.
\newblock \showarticletitle{Fairness definitions explained}. In
  \bibinfo{booktitle}{\emph{2018 IEEE/ACM International Workshop on Software
  Fairness (FairWare)}}. \bibinfo{publisher}{IEEE}, \bibinfo{pages}{1--7}.
\newblock


\bibitem[\protect\citeauthoryear{Wadsworth, Vera, and Piech}{Wadsworth
  et~al\mbox{.}}{2018}]%
        {wadsworth2018achieving}
\bibfield{author}{\bibinfo{person}{Christina Wadsworth},
  \bibinfo{person}{Francesca Vera}, {and} \bibinfo{person}{Chris Piech}.}
  \bibinfo{year}{2018}\natexlab{}.
\newblock \bibinfo{title}{Achieving fairness through adversarial learning: an
  application to recidivism prediction}.
\newblock
\newblock


\bibitem[\protect\citeauthoryear{Woodworth, Gunasekar, Ohannessian, and
  Srebro}{Woodworth et~al\mbox{.}}{2017}]%
        {woodworth2017learning}
\bibfield{author}{\bibinfo{person}{Blake Woodworth}, \bibinfo{person}{Suriya
  Gunasekar}, \bibinfo{person}{Mesrob~I Ohannessian}, {and}
  \bibinfo{person}{Nathan Srebro}.} \bibinfo{year}{2017}\natexlab{}.
\newblock \showarticletitle{Learning Non-Discriminatory Predictors}. In
  \bibinfo{booktitle}{\emph{Conference on Learning Theory}}.
  \bibinfo{pages}{1920--1953}.
\newblock


\bibitem[\protect\citeauthoryear{Xu, Yuan, Zhang, and Wu}{Xu
  et~al\mbox{.}}{2018}]%
        {xu2018fairgan}
\bibfield{author}{\bibinfo{person}{Depeng Xu}, \bibinfo{person}{Shuhan Yuan},
  \bibinfo{person}{Lu Zhang}, {and} \bibinfo{person}{Xintao Wu}.}
  \bibinfo{year}{2018}\natexlab{}.
\newblock \showarticletitle{Fairgan: Fairness-aware generative adversarial
  networks}. In \bibinfo{booktitle}{\emph{2018 IEEE International Conference on
  Big Data (Big Data)}}. \bibinfo{publisher}{IEEE}, \bibinfo{pages}{570--575}.
\newblock


\bibitem[\protect\citeauthoryear{Yao and Huang}{Yao and Huang}{2017}]%
        {yao2017new}
\bibfield{author}{\bibinfo{person}{Sirui Yao} {and} \bibinfo{person}{Bert
  Huang}.} \bibinfo{year}{2017}\natexlab{}.
\newblock \bibinfo{title}{New Fairness Metrics for Recommendation that Embrace
  Differences}.
\newblock
\newblock


\bibitem[\protect\citeauthoryear{Yeh and Lien}{Yeh and Lien}{2009}]%
        {yeh2009comparisons}
\bibfield{author}{\bibinfo{person}{I-Cheng Yeh} {and} \bibinfo{person}{Che-hui
  Lien}.} \bibinfo{year}{2009}\natexlab{}.
\newblock \showarticletitle{The comparisons of data mining techniques for the
  predictive accuracy of probability of default of credit card clients}.
\newblock \bibinfo{journal}{\emph{Expert Systems with Applications}}
  \bibinfo{volume}{36}, \bibinfo{number}{2} (\bibinfo{year}{2009}),
  \bibinfo{pages}{2473--2480}.
\newblock


\bibitem[\protect\citeauthoryear{Zafar, Valera, Gomez~Rodriguez, and
  Gummadi}{Zafar et~al\mbox{.}}{2017a}]%
        {zafar2017fairness}
\bibfield{author}{\bibinfo{person}{Muhammad~Bilal Zafar},
  \bibinfo{person}{Isabel Valera}, \bibinfo{person}{Manuel Gomez~Rodriguez},
  {and} \bibinfo{person}{Krishna~P Gummadi}.} \bibinfo{year}{2017}\natexlab{a}.
\newblock \showarticletitle{Fairness beyond disparate treatment \& disparate
  impact: Learning classification without disparate mistreatment}. In
  \bibinfo{booktitle}{\emph{Proceedings of the 26th International Conference on
  World Wide Web}}. \bibinfo{publisher}{International World Wide Web
  Conferences Steering Committee}, \bibinfo{pages}{1171--1180}.
\newblock


\bibitem[\protect\citeauthoryear{Zafar, Valera, Gomez~Rodriguez, and
  Gummadi}{Zafar et~al\mbox{.}}{2017b}]%
        {zafar2017AIAS}
\bibfield{author}{\bibinfo{person}{Muhammad~Bilal Zafar},
  \bibinfo{person}{Isabel Valera}, \bibinfo{person}{Manuel Gomez~Rodriguez},
  {and} \bibinfo{person}{Krishna~P Gummadi}.} \bibinfo{year}{2017}\natexlab{b}.
\newblock \showarticletitle{Fairness Constraints: Mechanisms for Fair
  Classification}. In \bibinfo{booktitle}{\emph{Artificial Intelligence and
  Statistics}}. \bibinfo{pages}{962--970}.
\newblock


\bibitem[\protect\citeauthoryear{Zemel, Wu, Swersky, Pitassi, and Dwork}{Zemel
  et~al\mbox{.}}{2013}]%
        {zemel2013learning}
\bibfield{author}{\bibinfo{person}{Rich Zemel}, \bibinfo{person}{Yu Wu},
  \bibinfo{person}{Kevin Swersky}, \bibinfo{person}{Toni Pitassi}, {and}
  \bibinfo{person}{Cynthia Dwork}.} \bibinfo{year}{2013}\natexlab{}.
\newblock \showarticletitle{Learning fair representations}. In
  \bibinfo{booktitle}{\emph{International Conference on Machine Learning}}.
  \bibinfo{pages}{325--333}.
\newblock


\bibitem[\protect\citeauthoryear{Zhang, Lemoine, and Mitchell}{Zhang
  et~al\mbox{.}}{2018}]%
        {zhang2018mitigating}
\bibfield{author}{\bibinfo{person}{Brian~Hu Zhang}, \bibinfo{person}{Blake
  Lemoine}, {and} \bibinfo{person}{Margaret Mitchell}.}
  \bibinfo{year}{2018}\natexlab{}.
\newblock \showarticletitle{Mitigating unwanted biases with adversarial
  learning}. In \bibinfo{booktitle}{\emph{Proceedings of the 2018 AAAI/ACM
  Conference on AI, Ethics, and Society}}. \bibinfo{publisher}{ACM},
  \bibinfo{pages}{335--340}.
\newblock


\bibitem[\protect\citeauthoryear{Zhang and Bareinboim}{Zhang and
  Bareinboim}{2018}]%
        {zhang2018fairness}
\bibfield{author}{\bibinfo{person}{Junzhe Zhang} {and} \bibinfo{person}{Elias
  Bareinboim}.} \bibinfo{year}{2018}\natexlab{}.
\newblock \showarticletitle{Fairness in decision-making—the causal
  explanation formula}. In \bibinfo{booktitle}{\emph{Thirty-Second AAAI
  Conference on Artificial Intelligence}}.
\newblock


\bibitem[\protect\citeauthoryear{Zhao, Wang, Yatskar, Cotterell, Ordonez, and
  Chang}{Zhao et~al\mbox{.}}{2019}]%
        {zhao2019gender}
\bibfield{author}{\bibinfo{person}{Jieyu Zhao}, \bibinfo{person}{Tianlu Wang},
  \bibinfo{person}{Mark Yatskar}, \bibinfo{person}{Ryan Cotterell},
  \bibinfo{person}{Vicente Ordonez}, {and} \bibinfo{person}{Kai-Wei Chang}.}
  \bibinfo{year}{2019}\natexlab{}.
\newblock \showarticletitle{Gender Bias in Contextualized Word Embeddings}. In
  \bibinfo{booktitle}{\emph{Proceedings of the 2019 Conference of the North
  American Chapter of the Association for Computational Linguistics: Human
  Language Technologies, Volume 1 (Long and Short Papers)}}.
  \bibinfo{pages}{629--634}.
\newblock


\bibitem[\protect\citeauthoryear{Zhao, Wang, Yatskar, Ordonez, and Chang}{Zhao
  et~al\mbox{.}}{2017}]%
        {zhao2017men}
\bibfield{author}{\bibinfo{person}{Jieyu Zhao}, \bibinfo{person}{Tianlu Wang},
  \bibinfo{person}{Mark Yatskar}, \bibinfo{person}{Vicente Ordonez}, {and}
  \bibinfo{person}{Kai-Wei Chang}.} \bibinfo{year}{2017}\natexlab{}.
\newblock \showarticletitle{Men Also Like Shopping: Reducing Gender Bias
  Amplification using Corpus-level Constraints}. In
  \bibinfo{booktitle}{\emph{Proceedings of the 2017 Conference on Empirical
  Methods in Natural Language Processing}}.
\newblock


\bibitem[\protect\citeauthoryear{Zhao, Wang, Yatskar, Ordonez, and Chang}{Zhao
  et~al\mbox{.}}{2018a}]%
        {zhao2018gender}
\bibfield{author}{\bibinfo{person}{Jieyu Zhao}, \bibinfo{person}{Tianlu Wang},
  \bibinfo{person}{Mark Yatskar}, \bibinfo{person}{Vicente Ordonez}, {and}
  \bibinfo{person}{Kai-Wei Chang}.} \bibinfo{year}{2018}\natexlab{a}.
\newblock \showarticletitle{Gender Bias in Coreference Resolution: Evaluation
  and Debiasing Methods}. In \bibinfo{booktitle}{\emph{Proceedings of the 2018
  Conference of the North American Chapter of the Association for Computational
  Linguistics: Human Language Technologies, Volume 2 (Short Papers)}}.
  \bibinfo{pages}{15--20}.
\newblock


\bibitem[\protect\citeauthoryear{Zhao, Zhou, Li, Wang, and Chang}{Zhao
  et~al\mbox{.}}{2018b}]%
        {zhao2018learning}
\bibfield{author}{\bibinfo{person}{Jieyu Zhao}, \bibinfo{person}{Yichao Zhou},
  \bibinfo{person}{Zeyu Li}, \bibinfo{person}{Wei Wang}, {and}
  \bibinfo{person}{Kai-Wei Chang}.} \bibinfo{year}{2018}\natexlab{b}.
\newblock \showarticletitle{Learning Gender-Neutral Word Embeddings}. In
  \bibinfo{booktitle}{\emph{Proceedings of the 2018 Conference on Empirical
  Methods in Natural Language Processing}}. \bibinfo{pages}{4847--4853}.
\newblock


\bibitem[\protect\citeauthoryear{Zimmer}{Zimmer}{1995}]%
        {zimmer1995emerging}
\bibfield{author}{\bibinfo{person}{Michael~J Zimmer}.}
  \bibinfo{year}{1995}\natexlab{}.
\newblock \showarticletitle{Emerging Uniform Structure of Disparate Treatment
  Discrimination Litigation}.
\newblock \bibinfo{journal}{\emph{Ga. L. Rev.}}  \bibinfo{volume}{30}
  (\bibinfo{year}{1995}), \bibinfo{pages}{563}.
\newblock


\bibitem[\protect\citeauthoryear{{\v{Z}}liobait{\.e}}{{\v{Z}}liobait{\.e}}{2017}]%
        {vzliobaite2017measuring}
\bibfield{author}{\bibinfo{person}{Indr{\.e} {\v{Z}}liobait{\.e}}.}
  \bibinfo{year}{2017}\natexlab{}.
\newblock \showarticletitle{Measuring discrimination in algorithmic decision
  making}.
\newblock \bibinfo{journal}{\emph{Data Mining and Knowledge Discovery}}
  \bibinfo{volume}{31}, \bibinfo{number}{4} (\bibinfo{year}{2017}),
  \bibinfo{pages}{1060--1089}.
\newblock


\end{thebibliography}

\newpage

\appendix
\section{Additional Measures for Algorithmic Fairness}
\label{S:app_measures}

Section \ref{S:measures} discussed the most prominent definitions and measures of algorithmic fairness.
This appendix reviews additional measures used in the literature.

\begin{enumerate}
\item\textbf{Overall accuracy equality} -- This requires similar accuracy across groups \citep{berk2018fairness}. This measure is mathematically formulated as follows:

\begin{equation} 
\label{eq7}
\left| P[Y=\hat{Y}|S=1]- P[Y=\hat{Y}|S\neq1] \right| \leq \varepsilon
\end{equation}

where $S$ represents the sensitive attribute (e.g., race and gender), $S=1$ is the privileged group and $S\neq1$ is the unprivileged group. $\hat{Y}=Y$ means that the prediction was correct. A lower value indicates better fairness. 
It should be noted that this measure does not guarantee \textit{equalized odds} or fair decisions (see \cite{Larson:2016:Online}).

\hfill

\item\textbf{Predictive parity} -- This requires that the \textit{positive predictive values} (PPVs) are similar across groups (meaning the probability of an individual with a positive prediction actually experiencing a positive outcome) \citep{chouldechova2017fair}. This measure is mathematically formulated as follows:

\begin{equation} 
\label{eq8}
\left| P[Y=1|S=1, \hat{Y}=1]-P[Y=1|S\neq1,\hat{Y}=1] \right| \leq \varepsilon
\end{equation}

Note that a lower value indicates better fairness. This measure uses the ground truth of the outcome, assuming that the outcome was achieved fairly. However, it has been shown to be incompatible with \textit{equalized odds} and \textit{equal opportunity} when prevalence differs across groups \citep{chouldechova2017fair,corbett2017algorithmic,corbett2018measure}.

\hfill

\item\textbf{Equal calibration} -- This requires that, for any predicted probability value, both groups will have similar positive predictive values (PPV represents the probability of an individual with a positive prediction actually experiencing a positive outcome) \citep{chouldechova2017fair,kleinberg2017inherent}. Note that this measure is similar to \textit{predictive parity} when the score value is binary (but does not guarantee predictive parity when the score is not binary) \citep{chouldechova2017fair}. This measure is mathematically formulated as follows:

\begin{equation} 
\label{eq9}
\left| P[Y=1|S=1,V=v]-P[Y=1|S\neq1, V=v] \right| \leq \varepsilon 
\end{equation}

where $V$ is the predicted probability value. Note that in some studies the definition of calibration requires that the PPV also be equal to $V$ \citep{kleinberg2017inherent,corbett2018measure}. A lower value indicates better fairness. Although in some cases equal calibration may be the desired measure, it has been shown that it is incompatible with \textit{equalized odds} \citep{pleiss2017fairness} and is insufficient to ensure accuracy or equitable decisions \citep{corbett2018measure}.
Moreover, it conflicts with \textit{balance for the positive class} and \textit{balance for the negative class} \citep{chouldechova2017fair,corbett2017algorithmic}.

\hfill

\item\textbf{Conditional statistical parity} -- Controlling for a limited set of "legitimate" features, an equal proportion of individuals is selected from each group \citep{corbett2017algorithmic}. This measure is mathematically formulated as follows:

\begin{equation} 
\label{eq10}
\left| P[\hat{Y}=1|S=1,L=l]-P[\hat{Y}=1|S\neq1 ,L=l] \right| \leq \varepsilon.   
\end{equation}

$L$ is a set of legitimate factors. A lower value indicates better fairness. Note that using this measure requires defining which features are legitimate, which is not a trivial task. It is not practical to find features that are entirely independent of the sensitive attributes.

\hfill

\item\textbf{Predictive equality} -- This requires false positive rates (FPRs; meaning the probability of an individual with a negative outcome to have a positive prediction) to be similar across groups \citep{corbett2017algorithmic}. This measure is mathematically formulated as follows:

\begin{equation} 
\label{eq11}
\left| P[\hat{Y}=1|S=1,Y=0]-P[\hat{Y}=1|S\neq1,Y=0] \right| \leq \varepsilon
\end{equation}

This measure requires the ground truth of the outcome, assuming that the outcome was achieved fairly. A lower value indicates better fairness. However, it considers only one type of error (as opposed to \textit{equalized odds}, for example, which require the equality of both FPRs and FNRs). As mentioned, following equality in terms of only one type of error will increase the disparity in terms of the other error \citep{pleiss2017fairness}.

\hfill

\item\textbf{Conditional use accuracy equality} -- This requires \textit{positive predictive values} (PPVs) and \textit{negative predictive values} (NPVs) to be similar across groups \citep{berk2018fairness}. NPV represents the probability of an individual with a negative prediction actually experiencing a negative outcome. PPV represents the probability of an individual with a positive prediction actually experiencing a positive outcome. This measure is mathematically formulated as follows:

\begin{equation} 
\label{eq12}
\begin{gathered}
\left| P[Y=1|S=1, \hat{Y}=1]-P[Y=1|S\neq1, \hat{Y}=1] \right|\leq \varepsilon 
\\
\land 
\\
\left| P[Y=0|S=1, \hat{Y}=0]-P[Y=0|S\neq1, \hat{Y}=0] \right|\leq \varepsilon
\end{gathered}
\end{equation}

This measure requires the ground truth of the outcome, assuming that the outcome was achieved fairly. A lower value indicates better fairness.
This measure considers more than one type of error. 
According to \cite{berk2018fairness}, following this measure does not guarantee \textit{equalized odds}.

\hfill

\item\textbf{Treatment equality} -- This requires an equal ratio of false negatives (FNs) and false positives (FPs) \citep{berk2018fairness}. The false negative cases are all of the cases that were predicted to be in the negative class when the actual outcome belongs to the positive class. The false positive cases are all of the cases that were predicted to be in the positive class when the actual outcome belongs to the negative class. This measure is mathematically formulated as follows:

\begin{equation} 
\label{eq13}
\left| FN_{S=1}/FP_{S=1}-FN_{S\neq1}/FP_{S\neq1} \right| \leq \varepsilon
\end{equation}

This measure requires the ground truth of the outcome, assuming that the outcome was achieved fairly. A lower value indicates better fairness. Note that according to \cite{berk2018fairness}, following this measure may harm the \textit{conditional use accuracy equality}.

\hfill

\item\textbf{Balance for the positive class} -- This requires an equal mean of predicted probabilities for individuals that experience a positive outcome \citep{kleinberg2017inherent}. This measure is mathematically formulated as follows:

\begin{equation} 
\label{eq14}
\left| E[V|Y=1,S=1]-E[V|Y=1,S\neq1] \right| \leq \varepsilon 
\end{equation}

where $V$ is the predicted probability value. A lower value indicates better fairness.
This measure was proven to be incompatible with \textit{equal calibration} \citep{kleinberg2017inherent}.

\hfill

\item\textbf{Balance for the negative class} -- This requires an equal mean of predicted probabilities for individuals that experience a negative outcome \citep{kleinberg2017inherent}. This measure is mathematically formulated as follows:

\begin{equation} 
\label{eq15}
\left| E[V|Y=0,S=1]-E[V|Y=0,S\neq1] \right| \leq \varepsilon
\end{equation}

where $V$ is the predicted probability value. A lower value indicates better fairness.
This measure was proven to be incompatible with \textit{equal calibration} \citep{kleinberg2017inherent}.

\hfill

\item\textbf{Fairness through unawareness} -- This requires that no sensitive attributes are explicitly used in the algorithm. The predicted outcomes are the same for candidates with the same attributes \citep{kusner2017counterfactual}. This measure is mathematically formulated as follows:

\begin{equation} 
\label{eq16}
X_{i}=X_{j}\rightarrow \hat{Y}_{i}=\hat{Y}_{j} 
\end{equation}

where $i$ and $j$ denote two individuals, and $X$ are the attributes describing an individual except for the sensitive attributes. This measure requires predictions to be the same for candidates with the same attributes. However, note that even when not considering the sensitive attributes, the model could still be biased through "proxies" or other causes such as sample or selection bias. Moreover, explicitly considering sensitive attributes is sometimes required, and excluding information can lead to discriminatory decisions \citep{corbett2018measure}.

\hfill

\item\textbf{Mutual information} -- This measures the mutual dependence between the sensitive feature and the predicted outcome \citep{kamishima2012enhancement}. This measure is mathematically formulated as follows:

\begin{equation} 
\label{eq17}
\sum\left (P(\hat{y},s)log(\frac{P(\hat{y},s)}{P(\hat{y})P(s)}) \right)\leq\varepsilon
\end{equation}

Note that this measure does not consider the actual outcomes. The lower the measure is, the lower the dependence between the sensitive attribute and the predictions; thus, lower values represent better fairness. One advantage of this measure is that it can consider binary, categorical or numerical predictions.

\hfill

\item\textbf{Mean difference} -- This measures the difference between the means of the predictions across groups \citep{vzliobaite2017measuring}. For example, \cite{louizos2016variational} use a variation of this measure that computes the difference between the means of the predicted probabilities across groups.

\begin{equation} 
\label{eq18}
\left| E[\hat{Y}|S=1]-E[\hat{Y}|S\neq1] \right|\leq \varepsilon
\end{equation}

Note that this measure does not consider the actual outcomes and that lower values indicate better fairness. One advantage of this measure is that it can consider binary, categorical or numerical predictions.

\end{enumerate}

\newpage
\noindent
Table \ref{tab:measures_appendix} presents a summary of the measures described in this appendix.

\fontsize{6}{12}\selectfont
\begin{longtable}
{|p{0.095\columnwidth}||p{0.03\columnwidth}|p{0.15\columnwidth}|p{0.065\columnwidth}|p{0.07\columnwidth}|p{0.08\columnwidth}|p{0.08\columnwidth}|p{0.08\columnwidth}|p{0.18\columnwidth}|} 
\caption{Additional Measures and Definitions for Algorithmic Fairness}{\label{tab:measures_appendix}}\\
\hline
\multicolumn{1}{|c||}{\textbf{Measure}} 
& \multicolumn{1}{c|}{\textbf{Paper}} 
& \multicolumn{1}{c|}{\textbf{Description}} & 
\multicolumn{1}{c|}{\textbf{Type}} 
& \textbf{Uses Actual Outcome} 
& \textbf{Uses Sensitive Attribute} 
& \textbf{Type of Actual Outcome} 
& \textbf{Type of Sensitive Attribute} 
& \multicolumn{1}{l|}{\textbf{Equivalent Notions}} 
\\ \hline \hline
\textbf{Overall Accuracy Equality} & \cite{berk2018fairness} 
& Requires similar accuracy across groups 
& Group 
& \cmark & \cmark & Binary & Binary
&
\\ \hline
\textbf{Predictive Parity} & \cite{chouldechova2017fair} 
& Requires that positive predictive values (PPVs) are similar across groups
& Group 
& \cmark & \cmark & Binary & Binary
& \vspace{-\baselineskip}\begin{itemize}[noitemsep,topsep=0pt,leftmargin=*,partopsep=0pt]
\item Equal positive predictive value (PPV) \citep{chouldechova2017fair}; 
\item Equal precision (see, for example, \cite{corbett2018measure})
\item Mathematically equal PPVs will induce equal false discovery rates (FDRs) (see \cite{verma2018fairness}). 
\vspace{-\baselineskip}\end{itemize}
\\ \hline
\textbf{Equal Calibration} & \cite{chouldechova2017fair}; \cite{kleinberg2017inherent} 
& Requires that for any predicted probability value, both groups will have similar positive predictive value (PPV)
& Group 
& \cmark & \cmark & Binary & Binary
& Similar to predictive parity when the score value is binary \citep{chouldechova2017fair} 
\\ \hline 
\textbf{Conditional Statistical Parity} & \cite{corbett2017algorithmic} 
& Controlling for a limited set of "legitimate" features, an equal proportion of individuals is selected from each group 
& Group 
& \xmark & \cmark & - & Binary
& Similar to the notion of \textit{fairness through unawareness} (see \cite{corbett2017algorithmic}) 
\\ \hline
\textbf{Predictive Equality} & \cite{corbett2017algorithmic} 
& Requires that false positive rates (FPRs) are similar across groups 
& Group 
& \cmark & \cmark & Binary & Binary
& \vspace{-\baselineskip}\begin{itemize}[noitemsep,topsep=0pt,leftmargin=*,partopsep=0pt]
\item False positive error rate balance \citep{chouldechova2017fair,verma2018fairness}
\item Equal false positive rates (FPRs) \citep{corbett2017algorithmic}
\item Mathematically equal FPRs will induce equal true negative rates (TNRs) (see \cite{verma2018fairness}) 
\vspace{-\baselineskip}\end{itemize}
\\ \hline
\textbf{Conditional Use Accuracy Equality} & \cite{berk2018fairness} 
& Requires that positive predictive values (PPVs) and negative predictive values (NPVs) are similar across groups.
& Group 
& \cmark & \cmark & Binary & Binary
&  
\\ \hline
\textbf{Treatment Equality} & \cite{berk2018fairness} 
& Requires equal ratio of false negatives (FNs) and false positives (FPs) 
& Group 
& \cmark & \cmark & Binary & Binary
&  
\\ \hline
\textbf{Balance for the Positive Class} & \cite{kleinberg2017inherent}
& Requires equal mean predicted probabilities for individuals that experience a positive outcome 
& Group 
& \cmark & \cmark & Binary & Binary
&  
\\ \hline
\textbf{Balance for the Negative Class} & \cite{kleinberg2017inherent}
& Requires equal mean predicted probabilities for individuals that experience a negative outcome 
& Group 
& \cmark & \cmark & Binary & Binary
&  
\\ \hline
\textbf{Fairness through Unawareness} & \cite{kusner2017counterfactual} 
& Requires that no sensitive attributes are explicitly used in the algorithm 
& Individual 
& \xmark & \xmark & - & -
& \vspace{-\baselineskip}\begin{itemize}[noitemsep,topsep=0pt,leftmargin=*,partopsep=0pt]
\item Fairness through blindness \citep{corbett2017algorithmic}
\item Anti-classification \citep{corbett2018measure}
\vspace{-\baselineskip}\end{itemize}
\\ \hline
\textbf{Mutual Information} & \cite{kamishima2012enhancement} 
& Measures mutual dependence between the sensitive feature and the predicted outcome 
& Group 
& \xmark & \cmark & - & Binary, Numerical, Categorical
& Related to prejudice index \citep{kamishima2012enhancement} 
\\ \hline
\textbf{Mean Difference} & \cite{vzliobaite2017measuring} 
& Measures the difference between the means of the targets across groups 
& Group 
& \xmark & \cmark & - & Binary, Numerical
& \vspace{-\baselineskip}\begin{itemize}[noitemsep,topsep=0pt,leftmargin=*,partopsep=0pt]
\item Discrimination probability \citep{louizos2016variational};
\item Similar to \textit{demographic parity} when the target is binary
\vspace{-\baselineskip}\end{itemize}
\\ \hline
\end{longtable}
\restoregeometry
\normalsize

\end{document}